\documentclass[%
reprint,
superscriptaddress, 
nofootinbib,
amsmath,amssymb,
aps,
pra,
floatfix,
]{revtex4-2}

\usepackage{graphicx}
\usepackage{dcolumn}
\usepackage{bm}
\usepackage{hyperref}[backref]

\usepackage[usenames,dvipsnames]{xcolor}
\hypersetup{
	colorlinks,
	allcolors=NavyBlue,
	linktoc=all
}

\usepackage{changes}
\usepackage{soul}

\usepackage{amsmath}
\usepackage{physics}
\usepackage{booktabs} 

\usepackage{float} 

\usepackage{microtype}

\usepackage{newtxtext}

\usepackage{url}

\usepackage{cases}

\newcommand{\matr}[1]{\mathbf{#1}} 

\usepackage{chemformula}

\usepackage{siunitx}

\begin{document}

\title{Quantum coherent microwave-optical transduction using high overtone bulk acoustic resonances}

\author{Terence Bl\'esin}
\email{terence.blesin@epfl.ch}
\affiliation{
	Institute of Physics, Swiss Federal Institute of Technology Lausanne (EPFL), Lausanne, Switzerland
}

\author{Hao Tian}
\affiliation{
	OxideMEMS lab, Purdue University, West Lafayette, IN, USA
}

\author{Sunil Bhave}
\affiliation{
	OxideMEMS lab, Purdue University, West Lafayette, IN, USA
}

\author{Tobias J. Kippenberg}
\email{tobias.kippenberg@epfl.ch}
\affiliation{
	Institute of Physics, Swiss Federal Institute of Technology Lausanne (EPFL), Lausanne, Switzerland
}



\begin{abstract}
	A device capable of converting single quanta of the microwave field to the optical domain is an outstanding endeavour in the context of quantum interconnects between distant superconducting qubits, but likewise can have applications in other fields, such as radio astronomy or, in the classical realm, microwave photonics.
	A variety of transduction approaches, based on optomechanical or electro-optical interactions, have been proposed and realized, yet the required vanishing added noises and an efficiency approaching unity, have not yet been attained.
	Here we present a new transduction scheme that could in theory satisfy the requirements for quantum coherent bidirectional transduction.
	Our scheme relies on an intermediary mechanical mode, a high overtone bulk acoustic resonance (HBAR), to coherently couple microwave and optical photons through the piezoelectric and strain-optical effects. Its efficiency results from the combination of integrated \ch{Si3N4} photonic circuits with ultra low loss sustaining high intracavity photon numbers with the highly efficient microwave to mechanical transduction offered by piezoelectrically coupled HBAR. 
	We develop a quantum theory for this multipartite system by first introducing a quantization method for the piezoelectric interaction between the microwave mode and the mechanical mode from first principles (which to our knowledge has not been presented in this form), and link the latter to the conventional Butterworth-Van Dyke model. The HBAR is subsequently coupled to a pair of hybridized optical modes from coupled optical ring cavities via the strain-optical effect. 
	We analyze the conversion capabilities of the proposed device using signal flow graphs, and demonstrate that near quantum coherent transduction is possible, with realistic experimental parameters. Combined with the high thermal conduction via the device bulk, heating effects are mitigated, and the approach does not require superconducting resonators that are susceptible to absorption of optical photons. 
\end{abstract}

\maketitle


\section{Introduction}

	Harnessing the effects of quantum mechanics is currently being investigating with respect to producing technologies with higher performance than their classical counter-parts in the domain of computing, sensing and communication, and proves to be a formidable challenge in terms of the required engineering.
	These improved performances of quantum technologies come at the expenses of overcoming difficult obstacles \cite{divincenzo2000physical}.
	Superconducting circuits are arguably among the most promising platforms for quantum computing owing the remarkably high nonlinearity of Josephson junctions \cite{devoret2013superconducting}.
	However, coupling between remote qubits has become an important consideration in the development of superconducting quantum computing on larger scales \cite{awschalom2019development}.
	While superconducting technologies enable consistently performing tasks in the quantum regime \cite{hofheinz2009synthesizing}, the GHz region of the electromagnetic spectrum they operate on is largely populated by thermal excitations, requiring cryogenic temperatures to preserve the quantum coherence of the logical units \cite{magnard_microwave_2020}.
	In contrast, optical signals have a carrier frequency in the 100's of THz region of the spectrum, with negligible thermal noise even at room temperature, and quantum states of light can be transported along optical fibers, with exceptionally low propagation loss ($\sim$ 0.5 dB/km) that have enabled optical fibers to become the major actor of high speed communication networks.
	One viable envisioned solution to the scalability issue is to process information with superconducting circuits at cryogenic temperature, and connect the different computing nodes with optical photons via optical fibers at room temperature \cite{awschalom2019development}.
	A quantum coherent transducer, the type of device addressed in this article, would establish a link between these two domains by coherently and bidirectionally converting single photons at microwave and optical frequencies with high efficiency and low added noise.
	In recent years, significant progress toward quantum coherent microwave-to-optical transduction has been made, primarily using cavity optomechanical, cavity electro-optical or piezooptomechanical conversion schemes, i.e. processes relying on radiation pressure coupling \cite{aspelmeyer2014cavity}, or the electro-optical equivalent \cite{tsang2010cavity}. The concept of these approaches is illustrated in the frequency domain on Fig. \ref{fig:FrequencyLandscape}. However, achieving conversion efficiencies approaching unity, which constitutes a stringent requirement to preserve quantum correlations of the converted signals \cite{caves1982quantum}, remains an outstanding engineering challenge, compounded by a variety of technical issues \cite{lauk2020perspectives}, including e.g. impedance matching, optical fiber-chip coupling, photorefractive effects, or even the vulnerability of superconductors to optical photons \cite{andrews2014bidirectional}.
	While there are many ways to overcome these challenges, it is nevertheless interesting to pursue the exploration  of alternative platforms for quantum coherent microwave-to-optical transduction, that alleviate some of these shortcomings. It is opportune to consider approaches based on wafer scale, high yield, ultra low loss integrated photonic circuits, in particular based on \ch{Si3N4}, which sustain large intracavity photon numbers, and have shown quality factors $Q>30 \times 10^{6}$ (i.e. propagation losses $\mathrm{<}$ 1dB/m) \cite{liu2020high, ji2017ultra}.
	
	The cavity optomechanical approach to microwave-optical conversion consists in parametrically coupling a mechanical mode to both microwave and optical cavities.
	Notable implementations include a \ch{Si3N4} membrane covered with \ch{Nb} in a Fabry-P\'erot optical cavity \cite{andrews2014bidirectional, higginbotham2018harnessing, mcgee2013mechanical} and \ch{Si} photonic crystal zipper cavities with \ch{Al} \cite{arnold2020converting}. This approach, although resorting to a low frequency mechanical mode that reduces the conversion bandwidth to the \si{\kilo\hertz} order, holds the record of measured efficiency of 40$\%$ for only 38 added noise photons \cite{higginbotham2018harnessing}. 
	
	For the cavity electro-optic approach, the microwave and optical fields are directly coupled via Pockels'effect, and the resulting Hamiltonian is equivalent to the optomechanical one \cite{tsang2010cavity}. Efficient conversion requires both electric and optical fields to overlap in a material with broken centrosymmetry, resulting in a three wave mixing process.
	Coupled racetracks resonators made from \ch{LiNbO3} fabricated on \ch{SiO2} \cite{holzgrafe2020cavity, soltani2017efficient} and sapphire \cite{mckenna2020cryogenic, xu2020bidirectional}, electrooptic polymer grown on \ch{Si} waveguides \cite{witmer2020silicon}, whispering gallery modes of bulk \ch{LiNbO3} crystals \cite{rueda2016efficient}, \ch{AlN} single \cite{fan2018superconducting} and coupled rings optical cavities \cite{fu2020ground}, as well as \ch{Ti} doped \ch{LiNbO3} phase modulators \cite{youssefi2020cryogenic} all fall into this category.
	These recent realizations have all relied on the use of coplanar waveguides or lumped elements LC resonators \cite{javerzac2016chip} to significantly increase the vacuum electric field strength, compared to bulk approaches \cite{ilchenko2003whispering, tsang2010cavity}. Yet, despite impressive advances, the approach is compounded by photorefractive effects limiting the available optical power and resonance frequency stability \cite{xu2021mitigating}.
	
	The third approach also makes use of an intermediary excitation, similarly to the first one, but the microwave is subsequently coupled to the mechanical oscillator using a piezoelectric material.
	The auxiliary mechanical mode is thus resonant with the microwave signal, which presents the advantage of exhibiting larger conversion bandwidths, on the order of \si{\mega\hertz}. Moreover, piezoelectric materials efficiently interconvert energy between electrical and mechanical forms, which earns them an ubiquitous place in MEMS and wireless communication technologies \cite{bhugra2017piezoelectric}.
	To complete the conversion process, the strain field of the mechanical mode couples phonons to the optical mode.
	This method has been implemented with suspended \ch{AlN} beams patterned as optomechanical crystals \cite{bochmann2013nanomechanical}, \ch{GaAs} photonic crystal nanobeams \cite{wu2020microwave, forsch2020microwave}, \ch{Si} photonic crystal excited by \ch{AlN} \cite{mirhosseini2020quantum}, \ch{LiNbO3} photonic crystal \cite{jiang2020efficient}, \ch{LiNbO3} thin film acoustic resonator \cite{shao2019microwave}, optical and acoustic whispering gallery modes of a \ch{LiNbO3} sphere \cite{yamazaki2020radio}, \ch{AlN} microdisk \cite{han2020cavity}.
	Impedance matching between the piezoelectrically excited acoustic excitations and mechanical modes hindering the conversion process, total conversion efficiencies up to $\mathcal{O}(10^{-5})$ have been achieved \cite{rueda2016efficient} so far.

	The transduction scheme presented in this paper falls in this last category. 
	In particular, we describe a modular interface designed for bidirectional conversion of travelling microwave and optical photons.
	It is inherently compatible with wafer-scale manufacturing and offers unique advantages for efficient conversion.
	Being based on ultra low loss \ch{Si3N4} photonic circuits fabricated using the damascene process, the transducer can sustain large intracavity photon numbers, and benefits from low fiber-chip insertion loss and reduced optical absorption \cite{liu2020high}.
	Bulk acoustic resonances are chosen to amplify the microwave signal rather than superconducting resonators in order not to suffer from Cooper pair breaking induced by scattered optical photons.
	In contrast to flexural modes of thin film nanomechanical and micromechanical resonators, the HBAR geometry allows for thermalization through the surrounding \ch{SiO2} cladding, and can additionally be cooled using \ch{^3He} buffer gas. The integration of the \ch{AlN} actuator on the cladding further allows for directly launching the acoustic wave in the HBAR cavity, circumventing the problem of acoustic impedance matching that can severely limit SAW-excited optomechanical crystals \cite{bochmann2013nanomechanical, forsch2020microwave}.
	Combined with the release process to increase the confinement of the HBAR mode in \ch{AlN}, a material with high piezoelectric coefficients \cite{oconnell_quantum_2010}, HBAR phonons fulfil their role as intermediaries between microwave and optical photons and allow for reaching high conversion efficiencies.
	
	Our manuscript is organized as follows.  We first describe the general idea of the proposed device in section \ref{section:description}.
	We then quantize the piezoelectric interaction in section \ref{section:piezoelectricity} in order to derive the correct coupling and decay rates that describe the conversion of the itinerant microwave photon to a phonon capable of interacting with the optical field, which to our knowledge has not been carried out in this form before. We furthermore explain how to experimentally measure these properties through microwave reflections.
	The basic theory of radiation pressure effects in the proposed device is exposed in section \ref{section:optomechanics}, together with simulations results justifying the expectation regarding the conversion efficiency exposed in section \ref{section:DeviceImplementation}.
	The formula for the conversion efficiency of the proposed scheme is derived in section \ref{section:TransductionTheory}, deriving the Langevin equations for the modes of the transducer from the total Hamiltonian, which deviate from prior treatments \cite{soltani2017efficient}. We introduce the use of signal flow graphs to interpret the dependence of the conversion efficiency on the system parameters, such as the cooperativities, in terms of feedback loops between the different modes.
	Finally, we give in section \ref{section:DeviceImplementation} an estimation of performance of the proposed scheme for realistic values of the physical parameters, obtained through simulations and confirmed by preliminary measurements, corresponding to devices currently under fabrication.

\section{Description of the device}  
\label{section:description}

	\begin{figure*}[htb]
		\centering
		\includegraphics[width=0.99\textwidth]{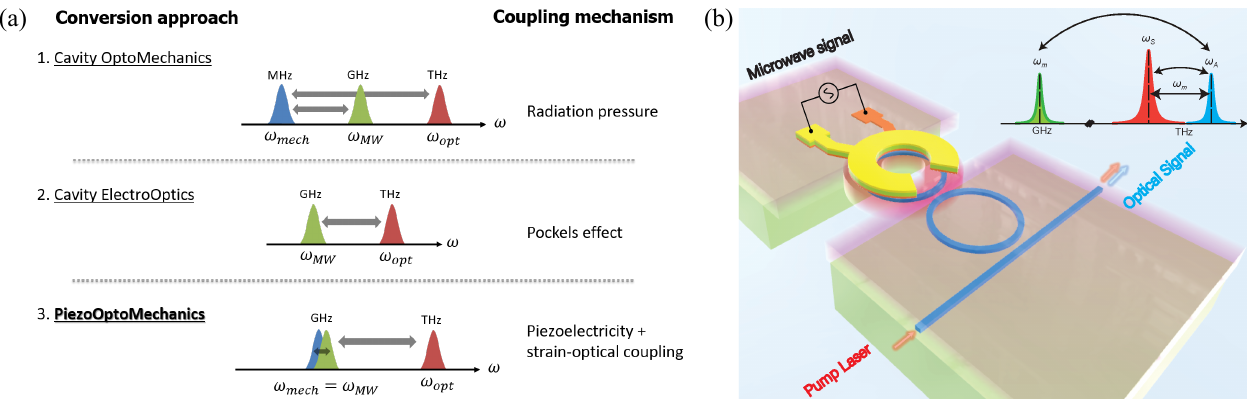}
		\caption{\textbf{Microwave-optical conversion schemes.}
				(a)
				1: Illustration of the schemes relying on the parametric coupling of both the optical cavity and MW cavity on a mechanical mode. 2: Schematic of the schemes relying on Pockels’effect to establish a direct link between MW and optics. 3: Schematic of the schemes relying on piezoelectricity to couple the MW to a mechanical mode acting on an optical cavity. This approach is the one described in this paper. Contrarily to the first approach, it exhibits a high conversion bandwidth as the mechanical frequency is matched to the electrical one. 
				(b)
				Artistic rendering of the proposed device.
				The electrodes (in yellow and orange) guide microwave photons to the piezoelectric material (in dark green) to be converted into HBAR phonons. The silicon substrate (in light green) is removed below the silicon oxide (in pink) HBAR cavity to reduce the mechanical mode volume. This increases the interaction with optical photons in the microring cavities (in blue).
				The inset in the top right of the figure indicates that a HBAR phonon annihilates with an optical photon of the symmetric supermode of the coupled rings to create an optical photon of the asymmetric supermode. 
				Conversely, an asymmetric supermode photon can be annihilated to create a phonon.}
		\label{fig:FrequencyLandscape}
		\label{fig:HBAR_3D}
	\end{figure*}
	
	This section presents the new type of devices discussed in this paper.
	The transducer proposed in this paper consists of a pair of coupled optical ring Si$_3$N$_4$ cavities in a SiO$_2$ cladding, that serves as resonator for a high overtone acoustic resonance (HBAR) addressed by a stack of electrodes and piezoelectric material, for example AlN.
	The aforementioned device, illustrated in Fig. \ref{fig:HBAR_3D}, is the result of recent efforts on the combination of AlN technology with silicon nitride integrated photonics that have led to the successful demonstration of on-chip control over soliton microcombs \cite{liu2020monolithic}, and the realization of an integrated CMOS-compatible optical isolator \cite{tian2020hybrid}.
	
	The evanescent coupling between cavity modes of identical silicon nitride micro-rings leads to the formation of effective supermodes, with a frequency splitting that can be controlled with the device geometry. This optical mode splitting is chosen to be equal to the frequency of the microwave signal to convert.
	The conversion proceeds by annihilating a pump photon and a microwave photon to create a photon in the other optical mode. 
	Direct transduction of single microwave photons thus requires pumping the symmetric optical supermode, which has a lower frequency than the asymmetric one, with an external laser. This induces a direct coupling between the mechanical mode and the unpopulated asymmetric optical supermode.
	By using rings with a small radius (circa 20 microns), the free spectral range of the optical cavity approaches \SI{1}{\tera\hertz}, and thus guarantees that the device operates in a region of the optical spectrum containing only the pair of optical supermodes  \cite{tikan2020emergent, zhang2019electronically}. 
	Furthermore, silicon nitride waveguides with very high intrinsic quality factors ($Q > 30 \times 10^6$) have already been demonstrated \cite{liu2020high}. This constitutes an important advantage as a slow decay rate of the optical photons in the cavity lowers the input power threshold for optimal conversion.   
	
	The interfacing of the microwave signal to the optics is achieved by a
	piezoelectric actuator (e.g. aluminium nitride) connected to the electrical circuit, whose photons should be upconverted to the optical domain. The voltage on the electrodes produces strain in the bottom layers that excites vertical acoustic waves in the stack. 
	Here we propose to etch part of the silicon layer beneath the silicon oxide cladding to control the size of the mechanical cavity, enabling control the mechanical mass and frequency of the high overtone bulk acoustic wave.
	This is a major difference with the modulators used in our previous work \cite{tian2020hybrid} as it gives a boost of almost two orders of magnitude in optomechanical coupling, and is expected to improve the quality factor of the HBAR resonator at low temperatures \cite{tian2020hybrid, liu2020monolithic, chu2017quantum, chu2018creation}.
	An important difference of the proposed device with previous converters is that there is no microwave resonator:
	it is the mechanical HBAR resonator itself that is overcoupled to the microwave signal and gives it a resonant enhancement.

\section{Quantum mechanical description of the piezoelectric interaction}
\label{section:piezoelectricity}

	\begin{figure*}
		\centering
		\includegraphics[width=0.95\textwidth]{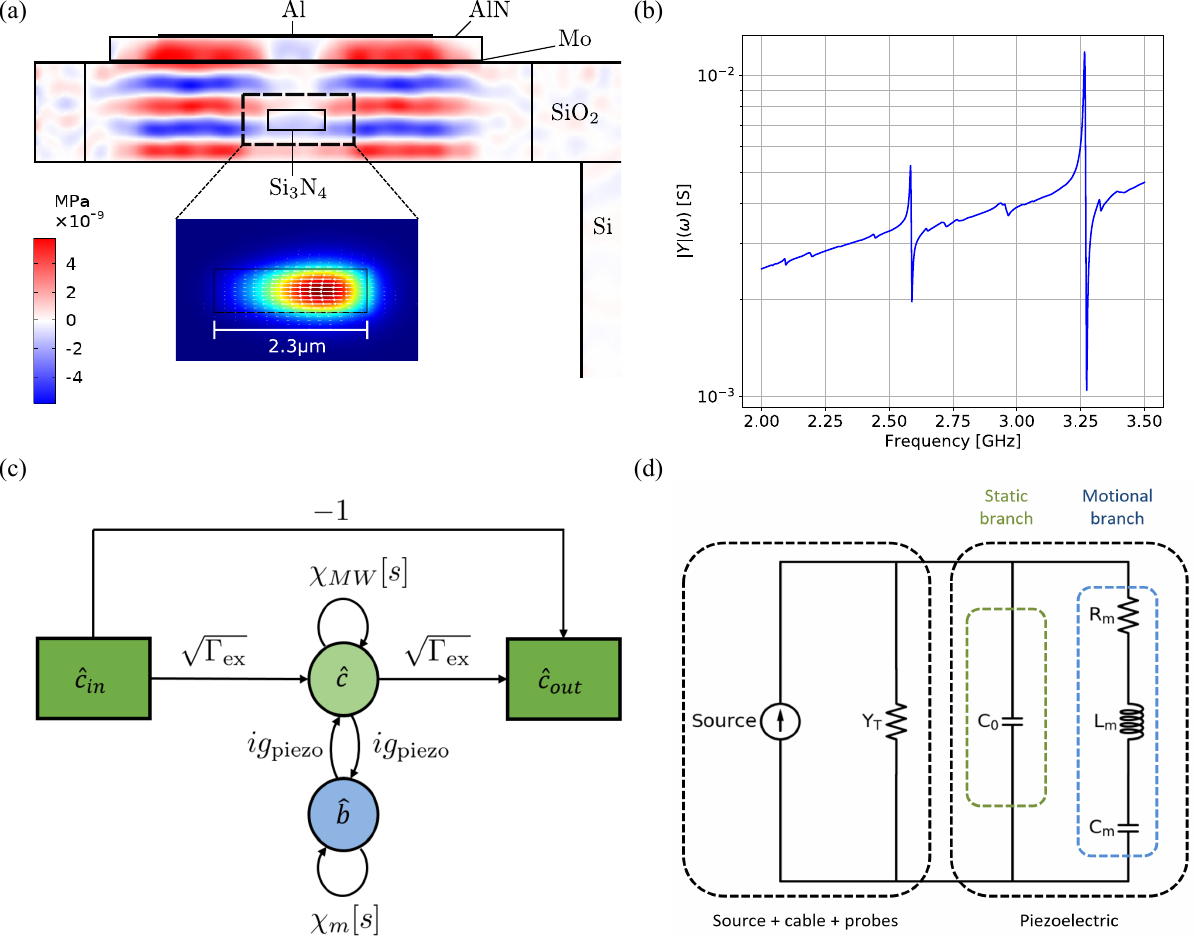}
		\caption{\textbf{Electromechanical conversion in the released HBAR device.}
				(a)
				Cross-section of the actual device showing the stress pattern of the HBAR mode excited by the electrodes at \SI{3.285}{\giga\hertz}.
				The inset shows the optical mode (norm of the electric field) in the optical ring waveguide. The small radius of the ring cavity of \SI{22.5}{\micro\meter} induces a shift of the mode from the waveguide center because of the strong bending potential.
				(b)
				Admittance curve of a released HBAR resonator obtained by finite element simulations. Each pair of resonance and antiresonance in the admittance $Y[\omega]$ corresponds to a mechanical resonance, around which the BVD model can be applied to model the interaction between the electric field and the mechanical motion. This curve shows two strong HBAR resonances around \SI{2.6}{\giga\hertz} and \SI{3.3}{\giga\hertz}. The exact frequency of the resonances and their optomechanical coupling rate depend on the position of the waveguide inside the cladding.
				(c)
				Signal flow chart to describe microwave reflection measurements.
				It also justifies that near resonance we can consider the travelling input microwave photon to be directly coupled to the mechanics.
				(d)
				Circuit diagram of the BVD model. It is used to model the behaviour of piezoelectric materials around a mechanical resonance.}
		\label{fig:MechanicalMode}
		\label{fig:Admittance}
		\label{fig:SignalFlowChart_S11}
		\label{fig:BVD}
	\end{figure*}
	
	This section describes how a mechanical mode in the bulk beneath the actuator exchanges energy with the electromagnetic field at microwave frequencies.
	In contrast to the description of piezoelectric materials usually given in terms of equivalent circuits \cite{zeuthen2018electrooptomechanical, arrangoiz2016engineering},
	we give an explanation in terms of scattering amplitudes by starting from the microscopic relations in the material and relying on energy conservation arguments.
	The advantage of this procedure is that it directly provides a link to quantum Langevin equations for solving the dynamics of the piezoelectric in the quantum regime, which has not been found in the existing literature to the best of the authors knowledge \cite{zeuthen2018electrooptomechanical, bhugra2017piezoelectric}. 
	We then discuss how the parameters appearing in this description can be related to the Butterworth-Van Dyke model (BVD), commonly used in the MEMS community \cite{bhugra2017piezoelectric} to characterize piezoelectric devices, and identify these parameters in microwave reflection measurements.
	
	\subsection{Equations of motion}
	
	The aim of this subsection is to derive a coupling rate between microwave photons and phonons in the HBAR cavity formed by the stack of electrodes, piezoelectric, waveguide and cladding.
	For this purpose, we start from the microscopic relations in piezoelectric materials \cite{bhugra2017piezoelectric}.
	Here we choose to use stress-charge formulation, which states that in these materials a strain will induce a polarization, and conversely an applied electric field will cause additional stress.
	\newline
	
	The stress $T_{ij}$ in the piezoelectric material contains two contributions, one from Hooke's law, via the stiffness coefficients $c_{ijkl}$, and one from the stress induced by an external electric field, via the piezoelectric stress coefficients $e_{kij}$ \cite{bhugra2017piezoelectric}
	\begin{equation}
		T_{ij} = \sum_{kl} c_{ijkl} S_{kl} - \sum_{k} e_{kij} E_{k}
	\label{eq:PiezoStress}
	\end{equation}
	Conversely, the electric displacement field in the piezoelectric material contains an additional contribution which originates from strain-induced charges
	\begin{equation}
		D_i = \sum_{k} \epsilon_{ik} E_k + \sum_{k l} e_{ikl} S_{kl}
		\label{eq:PiezoDisplacement}
	\end{equation}
	with $E_i$ as the electric field components, $S_{ij}$ as the strain components, and $\epsilon_{ij}$ as the elements of the permittivity tensor.
	
	Further study of the piezoelectric interaction requires the Hamiltonian density, which is derived from the Lagrangian $\mathcal{L}$ of the corresponding fields. The procedure to follow for deriving the Lagrangian from known equations of motion is  recalled in Appendix \ref{section:Lagrangian}.
	The Lagrangian density associated with mechanical deformations and consequent charges of a piezoelectric material can be derived as
	\begin{align}
		\mathcal{L} &= 
		\sum_{i} \frac{1}{2} \rho \dot{u}_i \dot{u}_i
		- \sum_{ijkl} \frac{1}{2} c_{ijkl} u_{i,j} u_{k,l} \nonumber \\
		& + \sum_{ij} \frac{1}{2} \epsilon_{ij} E_i E_j
		+ \sum_{ijk} e_{kij} E_k S_{ij}
	\end{align}
	with the two first term corresponding to the kinetic and elastic energy densities of the deformation, the third term to the energy density of the electric field and the last term coming from the piezoelectric interaction.
	Applying a Legendre transform to the variables $\{ u_i, E_i \}$ allows to recover the Hamiltonian density describing the acoustic and electric fields coupled in the piezoelectric
	\begin{align}
		\mathcal{H} &=
		\sum_{i} \frac{1}{2} \rho \dot{u}_i \dot{u}_i
		+ \sum_{ijkl} \frac{1}{2} c_{ijkl} u_{i,j} u_{k,l} \nonumber \\
		&- \sum_{ij} \frac{1}{2} \epsilon_{ij} E_i E_j
		- \sum_{ijk} e_{kij} E_k S_{ij}
	\end{align}
	The classical Hamiltonian is obtained by integrating this density over the whole space.
	The new term, only non-vanishing on the piezoelectric material, suggests the electromechanical interaction term on the electromagnetic field of the classical Hamiltonian to be
	\begin{equation}
		H_{\text{int}, \text{piezo}} = - \sum_{ikl} \iiint_{\text{piezo}} \left( e_{ikl} E_i S_{kl} \right)\ d\mathcal{V}
		\label{eq:PiezoHamiltonian}
	\end{equation}

	Having derived the interaction Hamiltonian from first principles, it is now possible to quantize the Hamiltonian by first quantizing the fields themselves.
	The mechanical displacement operator is defined as \cite{stroscio1996acoustic}
	\begin{equation}
		\hat{ \mathbf{u} }(\mathbf{r}, t) = \sqrt{\frac{\hbar }{2 m_{\text{eff}}\omega_m}} \mathbf{Q}(\mathbf{r}) \hat{b}(t) + h.c.
		\label{eq:DisplacementField}
	\end{equation}
	where $\hat{b}(t)$ is the annihilation operator describing the temporal evolution of the mechanical mode and will be discussed in the rest of the paper, $m_{\text{eff}}$ is the effective mass of the mode, related to the stiffness in the different materials encountered by the acoustic wave, and $\omega_m$ the eigenfrequency the mechanical mode.
	The mode function $\mathbf{Q}(\mathbf{r})$ is a solution of the spatial part of the acoustic wave equation for the conditions given by the geometry and materials of the device considered
	\begin{equation}
		- \omega_{m}^2 Q_i(\mathbf{r}) = \sum_{jkl} \frac{\partial}{\partial x_j} \left[ c_{ijkl}(\mathbf{r}) Q_{k, l}(\mathbf{r}) \right]
	\end{equation}
	This mode function, corresponding to an adimensional displacement function, is normalized such that 
	\begin{equation}
		\iiint_{\text{HBAR}} \rho(\mathbf{r}) \mathbf{Q}^*(\mathbf{r})\cdot\mathbf{Q}(\mathbf{r})\ d\mathcal{V} = m_{\text{eff}}
	\end{equation}
	to remain consistent with the definition given for the displacement operator.
	The prefactor in front of the spatial and temporal parts gives the displacement fluctuations of the zero-point of motion $u_{\text{ZPF}} = \sqrt{ \frac{\hbar}{2 m_{\text{eff}} \omega_m} }$.
	The strains are then obtained from the displacement operator $S_{ij} = \frac{1}{2} ( u_{i,j} + u_{j,i} )$.
	The antisymmetric part of the deformation can be overlooked as it is associated with local rigid rotations that have no implication in the current context: the strains are thus identified as $S_{ij} \sim u_{i,j}$ here \cite{ieee1984american}.
	
	Similarly the quantized form of the electric field relevant in this setting of piezoelectric interaction is \cite{cohen2017mecanique}
	\begin{equation}
		\hat{ \mathbf{E} }(\mathbf{r}, t) = - \sqrt{\frac{\hbar \omega_{\text{MW}}}{2\epsilon_{\text{eff}}}} \mathbf{F}(\mathbf{r}) \hat{c}(t) + h.c.
		\label{eq:MWfield}
	\end{equation}
	with an annihilation operator $\hat{c}(t)$ for the temporal evolution that will be formally eliminated from the description in a later section, an effective permeability seen by the mode $\epsilon_{\text{eff}}$ and a particular mode function $\mathbf{F}(\mathbf{r})$, solution of the spatial part of Helmholtz equation
	\begin{equation}
		\nabla \times \nabla \times \mathbf{F}(\mathbf{r}) = \frac{\epsilon(\mathbf{r})}{\epsilon_0} \frac{\omega_{\text{MW}}^2}{c^2} \mathbf{F}(\mathbf{r})
	\end{equation} 
	normalized such that
	\begin{equation}
		\iiint_{\text{whole space}} \frac{\epsilon(\mathbf{r})}{\epsilon_0} \mathbf{F}^*(\mathbf{r}) \cdot \mathbf{F}(\mathbf{r})\ d\mathcal{V} = \epsilon_{r, \text{eff}}
	\end{equation}
	to define the effective relative permittivity.
	
	The quantum mechanical Hamiltonian for the piezoelectric interaction between the acoustic and microwave fields is then obtained by inserting these quantized displacement (\ref{eq:DisplacementField}) and microwave (\ref{eq:MWfield}) fields in the interaction term \eqref{eq:PiezoHamiltonian}		
	\begin{widetext}
		\begin{equation}
			\hat{H}_{\text{int}, \text{piezo}} 
			= 
			\frac{- 1}{2} \hbar \sqrt{\omega_{\text{MW}}\omega_m}
			\underbrace{\left( \sum_{ikl} \iiint_{\text{piezo}} \sqrt{\frac{e_{ikl}^2}{ c_{\text{eff}}^E \epsilon_{\text{eff}}^T } }
				F_i(\mathbf{r}) \frac{1}{2} [Q_{k,l}(\mathbf{r}) + Q_{l, k}(\mathbf{r})] \ d\mathcal{V} \right)}_{\sqrt{k_{\text{eff}}^2}}
			\left( \hat{b} + \hat{b}^{\dagger} \right) \left( \hat{c} + \hat{c}^{\dagger} \right)
		\end{equation}
	\end{widetext}
	
	When defining an effective stiffness $c_{\text{eff}}^{E} = m_{\text{eff}} \omega_m^{2}$ for the mechanical mode and an effective permittivity for the corresponding microwave mode $\epsilon^{T}$ at constant strain, we notice the emergence of the \emph{electromechanical coupling factor} as defined by the MEMS community \cite{ieee1984american}
	\begin{equation}
		K^2 = \frac{e^2}{c^{E} \epsilon^{T}}
	\end{equation}
	The integral over the piezoelectric volume of this quantity then gives the effective electromechanical coupling factor $\sqrt{k_{\text{eff}}^2}$, a quantity easily extracted from the BVD model as explained in the next section.
	Defining a piezoelectric electromechanical coupling rate
	\begin{equation}
		g_{\text{EM}} 
		= \frac{1}{2} \sqrt{k_{\text{eff}}^2} \sqrt{\omega_{m} \omega_{\text{MW}}}
		\label{eq:PiezoCouplingRate}
	\end{equation}
	the quantized piezoelectric interaction Hamiltonian finally takes the form
	\begin{equation}
		\hat{H}_{\text{int}, \mathrm{piezo}} = 
		- \hbar g_{\text{EM}}
		\left( \hat{b} + \hat{b}^{\dagger} \right) \left( \hat{c} + \hat{c}^{\dagger} \right)
	\end{equation}	
	Importantly, the value of this coupling rate is independent of the amplitude of the strain field and electric field, being only determined by the material properties and geometry of the device, as expected for linear piezoelectrics.
	We note that this coupling rate has the same form as the one derived by E. Zeuthen \textit{et al.}, although the derivation is based on a different model \cite{zeuthen2018electrooptomechanical, wu2020microwave}.
	
	\subsection{Butterworth-Van Dyke model}
	
	For experimental characterization of piezoelectrics, it is common to resort to an equivalent circuit, which was conceived by Butterworth and Van Dyke \cite{bhugra2017piezoelectric}.
	While later modifications were brought to better fit experimental data, the essence of the model lies in the parallel coupling between a branch containing the behaviour of the electrodes in absence of piezoelectric effect with a branch representing the motion of the mechanical degree of freedom, up to some voltage conversion factor.
	The static capacitance $C_0$ takes into account the usual capacitive behaviour of the electrodes, from the energy stored via the permittivity of the material without the piezoelectric effect.
	The motional branch connected in parallel arises from the supplementary charges created on the electrodes by the mechanical deformation via the piezoelectric effect. 
	The motional capacitance $C_m$ accounts for the potential mechanical energy while the motional inductance $L_m$ accounts for the kinetic mechanical energy. Finally the dissipation of acoustic waves is modelled by a motional resistance $R_m$.
	This equivalent circuit is illustrated in Fig. \ref{fig:BVD}.
	Piezoelectric devices are usually characterized by looking at their admittance, which is ideally represented by the BVD model admittance
	\begin{equation}
		Y(\omega) = i \omega C_0 + \frac{1}{L_m} \frac{i\omega}{-\omega^2 + i \omega \gamma_0 + \omega_m^2}
	\end{equation}
	near a resonance. $\omega_m$ and $\gamma_0$ are respectively the resonant frequency and linewidth of the electrically probed mechanical resonance.
	A few parameters determining the behaviour of the device and quality of the mechanical resonance can be extracted from the curve of the admittance as a function of the frequency.
	Each mechanical mode will exhibit a peak at what is called a series resonance frequency $\omega_s$ and a dip at what is called a parallel resonance frequency $\omega_p$.
	Away from the resonance the device behaves as a capacitor of value $C_0$.
	The bandwidth of the conversion between electromagnetic energy and mechanical energy is described by the effective electromechanical coupling factor
	\begin{equation}
		k_{\text{eff}}^2 = \frac{\omega_p^2 - \omega_s^2}{\omega_p^2}
	\end{equation}
	which also fixes the value of the electromechanical coupling rate.
	This coupling factor then indicates the scale of the motional capacitance compared to the static effect
	\begin{equation}
		C_m = k_{\text{eff}}^2 C_0
	\end{equation}
	The last parameters $L_m$ and $R_m$ are then extracted from the frequency $\omega_m$ and linewidth $\gamma_0$ of the mechanical resonance.
	\footnote{Comparing the microwave reflection spectrum and the admittance shows that the mechanical resonance is in the middle of the series and parallel resonances $\omega_m = \frac{\omega_s + \omega_p}{2}$.}
	\begin{align}
		L_m &= \frac{1}{\omega_m^2 C_m} \\
		R_m &= L_m \gamma_0
	\end{align}
	
	Fig. \ref{fig:Admittance} shows the admittance curve of the HBAR resonator considered here. This curve was obtained by finite elements simulations.
	Each mechanical resonance features a succession of resonance and antiresonance on the admittance. The height of the resonance indicates the strength of the coupling to the microwave mode.
	While the simulated curve exhibits several resonant features, two particular frequencies, corresponding to the silicon oxide HBAR modes, are standing out near \SI{2.6}{GHz} and \SI{3.3}{\giga\hertz}.
	The latter one is the one considered for the numerical evaluations presented in the rest of the paper as its electromechanical coupling factor takes a higher value. \\

	\subsection{Link to MW reflection measurements}
	
	This section attempts to reconcile the engineers picture, primarily using the notions of voltages, currents and immitance, with the physicists picture, relying on coupling rates for an Hamiltonian formulation, by showing how reflection measurements performed in  laboratory are linked to the ladder operators description, which holds a central place in second quantization.
	The admittance curve fitted by the BVD model as described in the previous section is generally obtained via microwave reflection measurements $S_{11}[\omega]$. These measurements are also easily described in terms of annihilation operators.
	A single microwave mode interacting with a single mechanical mode via the piezoelectric interaction 
	\begin{equation}
		\hat{H}_{\text{piezo}} = 
		- \hbar g_{\text{EM}}
		\left( \hat{b} + \hat{b}^{\dagger} \right) \left( \hat{c} + \hat{c}^{\dagger} \right)
	\end{equation}
	are evolving according to the coupled modes equations
	\begin{align}
		\frac{d}{dt} \hat{b} = 
		&-i \omega_{m} \hat{b}  
		- \frac{\gamma_0}{2} \hat{b} \nonumber \\
		&+ i g_{\text{EM}} \left( \hat{c} + \hat{c}^{\dagger} \right) 
		+ \sqrt{\gamma_{0}} \hat{f}_{m} \\
		\frac{d}{dt} \hat{c} = 
		&-i \omega_{\text{MW}} \hat{c}  
		- \frac{\Gamma}{2} \hat{c} \nonumber \\
		&+ i g_{\text{EM}} \left( \hat{b} + \hat{b}^{\dagger} \right)
		+ \sqrt{\Gamma_{0}} \hat{f}_{\text{MW}}
		+ \sqrt{\Gamma_{\text{ex}}} \hat{c}_{\text{in}}
	\end{align}
	
	Taking Laplace's transform 
	\footnote{Defined here and in the rest of the article such that $f(t) \rightarrow f[s] = \int_{0}^{\infty} e^{-s t} f(t) dt $}
	and defining susceptibilities for the microwave field
	\footnote{
		$s$ is the complex frequency from Laplace transform. Its imaginary part is related to the frequency with the physicist convention as $\text{Im}(s) = -\omega$. This convention is used in the rest of the paper to go from Laplace transform to the frequency domain.
	}
	
	\begin{equation}
		\chi_{\text{MW}}[s] = \frac{1}{s + i\omega_{\text{MW}} + \frac{\Gamma}{2}}
	\end{equation}
	and for the strain field
	\begin{equation}
		\chi_{m}[s] = \frac{1}{s + i\omega_{m} + \frac{\gamma_0}{2}}
	\end{equation}
	the microwave reflection coefficient $S_{11}[\omega]$ is readily obtained from the signal flow graph on Fig. \ref{fig:SignalFlowChart_S11} using Mason's gain formula \cite{mason1956feedback}
	\begin{equation}
		S_{11}[s] = \frac{\hat{c}_{\text{out}}}{\hat{c}_{\text{in}}}
		= -1 + \Gamma_{\text{ex}} \frac{ \chi_{\text{MW}}[s] }{1 + g_{\text{EM}}^2 \chi_{\text{MW}}[s]\chi_m[s] }
	\end{equation} \\
	
	Finally, we wish to justify the form of the equation for the mechanics that will be used in the section on the conversion efficiency of the transducer, where the mechanical mode is considered to be directly coupled to the input microwave photon, without having to go through an intermediary microwave mode with an equation of the type
	\begin{equation}
		\frac{d}{dt} \hat{b} = 
		-i \omega_{m} \hat{b}  
		- \frac{\gamma_m}{2} \hat{b}
		+ \sqrt{\gamma_{\text{piezo}}} \hat{f}_{\text{piezo}}
		+ \sqrt{\gamma_{\text{ex}}} \hat{c}_{\text{in}}
	\end{equation}
	to which the optomechanical interaction will be added.
	The main point is that the internal dynamics of the microwave mode can be hidden to the optics, which will effectively see the mechanics as if it was directly pumped by the incoming microwave photons.
	This is done using the microwave mode susceptibility to express it as a function of the other terms in the coupled modes of the piezoelectric
	\begin{align}
		\hat{c}[s] &= i \chi_{\text{MW}}[s] g_{\text{EM}} \hat{b}[s] \nonumber \\
		&+ \chi_{\text{MW}}[s] \sqrt{\Gamma_{0}} \hat{f}_{\text{MW}}[s]
		+ \chi_{\text{MW}}[s] \sqrt{\Gamma_{\text{ex}}} \hat{c}_{\text{in}}[s]
	\end{align}
	which fixes the coefficients to use in the desired equation from the original coefficients of the separate mechanical and microwave systems:
	\begin{align}
		\begin{cases}
			\frac{\gamma_m}{2} = \frac{\gamma_0}{2} + \chi_{\text{MW}}[s] g_{\text{EM}}^2 \\
			\sqrt{\gamma_{\text{noise}}} \hat{f}_{\text{piezo}} =
			\sqrt{\gamma_{0}} \hat{f}_{m} 
			+ i g_{\text{EM}} \chi_{\text{MW}}[s] \sqrt{\Gamma_{0}} \hat{f}_{\text{MW}} \\
			\sqrt{\gamma_{\text{ex}}} =
			i g_{\text{EM}} \chi_{\text{MW}}[s] \sqrt{\Gamma_{\text{ex}}}
		\end{cases}
	\end{align}
	Since the electrodes are directly connected to the source and no inductive element is added in the circuit, we can assume that the coupling rate from the source to the microwave mode is directly given by the charging time of the static capacitance $C_0$ by the characteristic impedance $Z_0 = \SI{50}{\ohm}$
	\begin{equation}
		\Gamma_{\text{ex}} = \frac{1}{Z_0 C_0}
	\end{equation}
	Similarly the intrinsic decay rate of the microwave mode is given by the discharge time of the static capacitance in the parallel resistance of the modified BVD circuit shown in Fig. \ref{fig:BVD}
	\begin{equation}
		\Gamma_0 = \frac{1}{R_0 C_0}
	\end{equation}
	
	Simulations and preliminary experiments show that $C_0 \approx \SI{200}{\femto\farad}$ and $R_0 \approx \SI{10}{\kilo\ohm}$.
	For these values the microwave mode has a total linewidth $\Gamma = \Gamma_0 + \Gamma_{\text{ex}} \approx \SI{15}{\giga\hertz}$ and is strongly overcoupled to the source, with $\frac{\Gamma_{\text{ex}}}{\Gamma} > 95\%$.
	Under these conditions, the effective linewidth of the mechanical mode addressed by the microwave source is
	\begin{equation}
		\gamma_m \approx \gamma_0 + 4 \frac{g_{\text{EM}}^2}{\Gamma}
	\end{equation}
	and the microwave photon to phonon coupling rate is
	\begin{equation}
		|\gamma_{\text{ex}}| 
		\approx
		\frac{4 g_{\text{EM}}^2}{\Gamma} 
		\frac{\Gamma_{\text{ex}}}{\Gamma}
	\end{equation}
	The large linewidth and strong overcoupling of the microwave mode explain why mechanical modes of piezoelectric resonators can be directly probed with microwave reflection measurements, and why no mode splitting is observed despite having an electromechanical coupling rate greater than the mechanical linewidth.

\section{Optomechanical interaction}
\label{section:optomechanics}
	
	In this section we move to the optical side of the device, which is interfaced to the microwave part via the mechanics.
	The optomechanical interaction essentially expresses the fact that the optical energy can be transferred to a mechanical degree of freedom, and vice-versa.
	For the proposed device, composed of a silicon nitride waveguide in silicon oxide cladding, this phenomenon finds its root in two effects:
	either the optical field can act on the mechanical field via radiation pressure and the latter reacts via moving boundaries, or the optical field acts on the mechanical field via electrostriction and the latter reacts by photoelasticity.
	
	\subsection{Photoelastic effect}
	
	Strains in the material are changing the potential seen by the electrons, which change the permittivity of the medium.
	This directly translate to a change in refractive index, and this effect is called photoelasticity.
	
	Mathematically, photoelasticity is easily expressed as a change in the impermeability $\eta = \frac{1}{\epsilon_r}$
	\begin{equation}
	\Delta \eta_{ij} = \sum_{kl} p_{ijkl} S_{kl}
	\end{equation}
	The modulation of impermeability obviously corresponds to a modulation in relative permittivity
	\begin{equation}
	(\Delta \epsilon_r)_{ij} 
	= \sum_{k l} - \epsilon_{r, ik} \epsilon_{r, kl} (\Delta \eta)_{lj}
	\end{equation}
	A change in permittivity corresponds to a variation of the energy stored in the electromagnetic field, and thus a change in the optical frequency given by Bethe-Schwinger formula for optical cavity perturbation \cite{betheschwinger1943}
	\begin{equation}
	\Delta \omega = \frac{-\omega_0}{2} \frac{\iiint \expval{\Delta\epsilon}{E}d\mathcal{V}}{\iiint \expval{\epsilon}{E}d\mathcal{V}}
	\end{equation}
	The optomechanical coupling rate for a single photon depends on the cavity frequency shifts and the zero point fluctuations motion of the quantity causing the modulation, here the displacement in the waveguide and cladding $g_0 = \frac{\partial \omega}{\partial S} u_{\text{ZPF}}$.\footnote{Here we used $S_{\text{ZPF}} = u_{\text{ZPF}}$.}
	The contribution to the optomechanical single photon coupling strength from photoelasticity is thus
	\begin{equation}
	g_{0, \text{OM,PE}} =
	- \frac{\omega_0}{2} \frac{\expval{\mathbf{E}{\frac{\partial \epsilon}{\partial S}}{\mathbf{E}}}}{\int \mathbf{E}\cdot\mathbf{D}\  d\mathcal{V}} u_{\text{ZPF}}
	\end{equation}

	\subsection{Moving boundaries effect}
	
	The deformation of the waveguide changes the confinement of the optical mode in the different media, which in return modifies its effective refractive index.
	The associated change in frequency is again computed using Bethe-Schwinger formula for cavity perturbation.
	\begin{equation}
	g_{0, \text{OM,MB}} =
	- \frac{\omega_0}{2} \frac{\oint (\mathbf{S}_{\text{ZPF}} \cdot \bar{\mathbf{n}}) (\Delta\epsilon \mathbf{E}_{||}^2 - \Delta\epsilon^{-1}\mathbf{D}_{\perp}^2) d\mathcal{S}}{\int \mathbf{E}\cdot\mathbf{D} d\mathcal{V}}
	\end{equation}
	Here $\bar{\mathbf{n}}$ denotes the normal vector of the interface between Si$_3$N$_4$ and SiO$_2$, $\Delta \epsilon = \epsilon_{\ch{Si3N4}} - \epsilon_{\ch{SiO2}}$ is the permittivity difference between supporting media, $\Delta \epsilon^{-1} = \epsilon_{\ch{Si3N4}}^{-1} - \epsilon_{\ch{SiO2}}^{-1}$ is the impermeability difference between the supporting media, $\mathbf{E}_{||}$ is the electric field component parallel to the interface and $\mathbf{D}_{\perp}$ is the electric displacement component perpendicular to the interface.
	We note that this effect can cause a frequency shift with a sign different from photoelasticity, and thus can sometimes reduce the single photon coupling strength.
	
	\subsection{Estimation of the single photon coupling strength}
	
	The mechanical behaviour of the undercut SiO$_2$ HBAR resonator was simulated via finite element methods using the COMSOL software, including the piezoelectric effect.
	The simulations were used to obtain the admittance curve, from which the resonances could be determined, and to compute the single photon optomechanical coupling strength, including both photoelasticity and moving boundaries effects described in the previous subsections.
	As the optical waveguides are made of silicon nitride and silicon oxide in the amorphous phase, symmetry arguments ensure that the form of the photoelastic coefficients tensor is of the kind
	\begin{equation}
	\mathbf{p}_{\text{amorphous}} =
	\left(\begin{matrix}p_{11} & p_{12} & p_{12} & 0 & 0 & 0\\p_{12} & p_{11} & p_{12} & 0 & 0 & 0\\p_{12} & p_{12} & p_{11} & 0 & 0 & 0\\0 & 0 & 0 & p_{44} & 0 & 0\\0 & 0 & 0 & 0 & p_{44} & 0\\0 & 0 & 0 & 0 & 0 & p_{44}\end{matrix}\right)
	\end{equation}
	with $p_{44} = \frac{1}{2}(p_{11} - p_{12})$.
	The values of the photoelastic coefficients of silicon nitride have only been reported by few  authors. Here we chose the latest values reported in literature: $p_{11} = 0.239$ \cite{gyger2020observation} and $p_{12} = 0.047$ \cite{capelle2017polarimetric}, giving $p_{44} = 0.096$.
	The coefficients used for the silicon oxide cladding, which contains around $30\%$ of the TE optical mode volume were:
	$p_{11} = 0.121$, $p_{12} = 0.270$ and $p_{44} = - 0.0745$ \cite{huang2003stress}.
	
	The inset in Fig. \ref{fig:MechanicalMode}\textcolor{NavyBlue}{a}) shows the profile of the norm of the electric field of the optical mode on which the device will operate, while the core of Fig. \ref{fig:MechanicalMode}\textcolor{NavyBlue}{a}) shows the pattern of the total stress along the vertical direction of an HBAR resonance.
	The \SI{900}{\nano\meter} thick \ch{Si3N4} waveguide is buried in \SI{4}{\micro\meter} of \ch{SiO2} cladding, with a \SI{1}{\micro\meter} thick \ch{AlN} actuator.
	As the acoustic impedance of the Si$_3$N$_4$ waveguide did not match the acoustic impedance of the rest of the stack, its position had to be optimized to obtain high optomechanical coupling simultaneously with high electromechanical coupling.
	In the end the position of the waveguide could be chosen such that the strong acoustic resonance around \SI{3.285}{\giga\hertz} shows a single photon optomechanical coupling rate $\frac{g_0}{2\pi} \approx \SI{400}{\hertz}$. \\

\section{Microwave-to-optical transduction}
\label{section:TransductionTheory}

	\begin{figure*}[htb]
		\centering
		\includegraphics[width=0.95\textwidth]{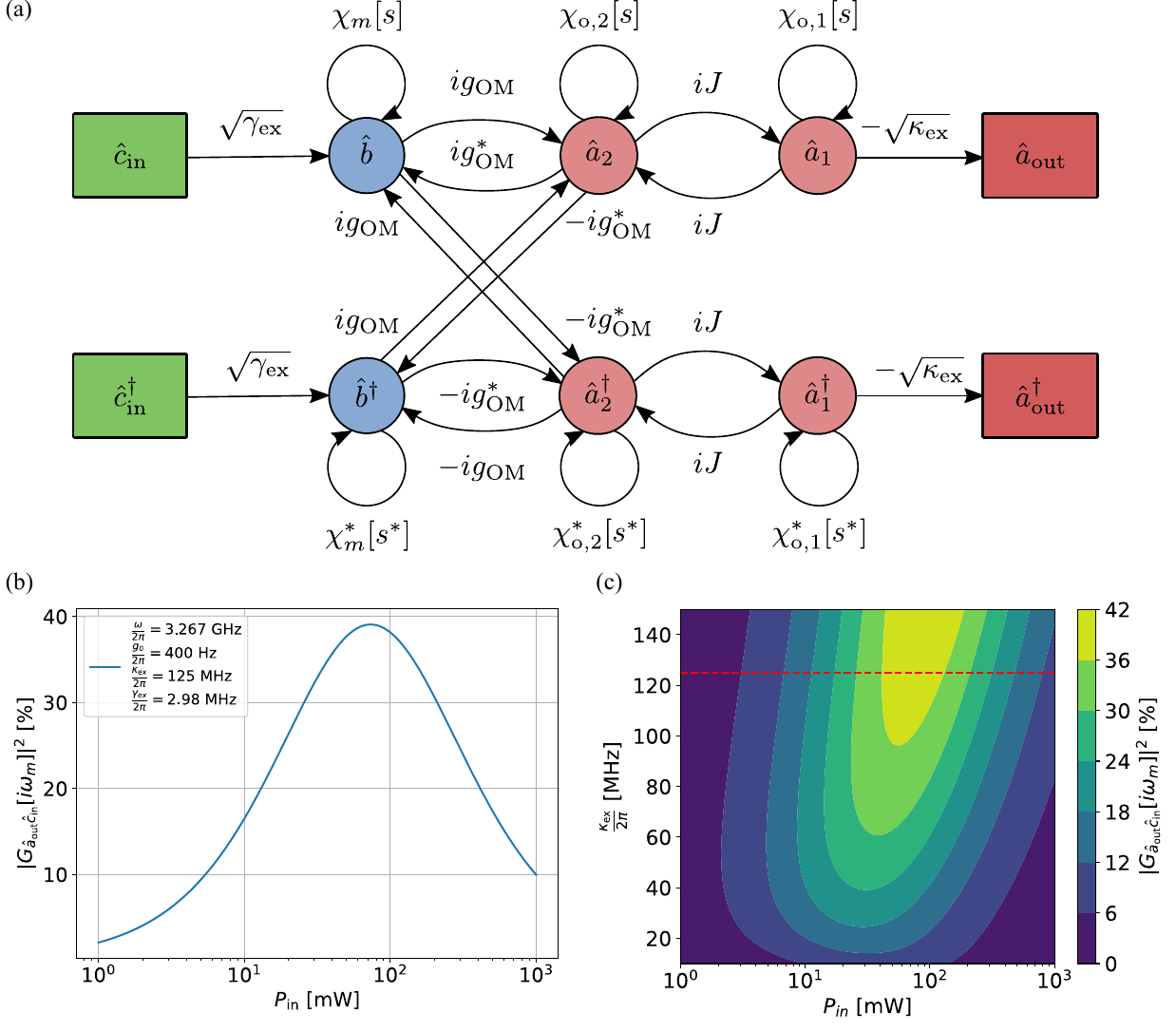}
		\caption{\textbf{Microwave-optical conversion efficiency.}
				(a)
				Signal flow chart describing the MW-optical conversion process. The squares represent the input and output modes, while the circles represent the internal modes of the converter. The arrows represent the coupling between the modes and are weighted by the value of the coupling rate between the modes. The transfer functions between two different modes can be computed with this diagram using Mason's gain formula.
				(b)
				Theoretical conversion efficiency from the microwave port to the transmitted optical asymmetric supermode as a function of the input pump laser power, obtained from eq.(\ref{eq:RWA_efficiency}). The parameters used to draw this curve are indicated in the textbox as well as in table (\ref{tab:CurrentParameters}), and correspond to the dashed red line in Fig. \ref{fig:ConversionEfficiency_sweeps}\textcolor{NavyBlue}{c}).
				(c)
				Microwave to optical conversion efficiency as a function of the input pump laser power and the coupling rate from the bus waveguide to the first ring cavity. The parameters used to draw this curve are indicated in table (\ref{tab:CurrentParameters}).}
		\label{fig:SignalFlowChart}
		\label{fig:ConversionEfficiency}
		\label{fig:ConversionEfficiency_sweeps}
	\end{figure*}
	
	In this section,  we finally derive the expression of the conversion efficiency between the input microwave photon and the output upconverted optical photon. Details on the derivation can be found in the Appendices (\ref{appendix:Hamiltonian}) and (\ref{appendix:TransferFunctions}).
	
	Here we consider a pair of coupled optical modes, of which only one is accessed by a bus waveguide to receive and transmit power, while the other interacts with the mechanical mode, accessed by a microwave mode. This device is depicted on Fig. \ref{fig:HBAR_3D}.
	The input-output relation for the optical transmission through the bus waveguide coupled to the first microring cavity is
	\begin{equation}
		\hat{a}_{\text{out}} = \hat{a}_{\text{in}} - \sqrt{\kappa_{\text{ex}}} \hat{a}_1
	\end{equation}
	while the input-output relation for the microwave reflection from the HBAR is
	\begin{equation}
		\hat{c}_{\text{out}} = -\hat{c}_{\text{in}} + \sqrt{\gamma_{\text{ex}}} \hat{b}
	\end{equation}
	The Hamiltonian describing the internal dynamics of this system is given by
	\begin{equation}
		\label{eq:Hamiltonian}
		\begin{split}
		\hat{H} =\ &
		\hbar \omega_1 \hat{a}_1^{\dagger}\hat{a}_1
		+ \hbar \omega_2 \hat{a}_2^{\dagger}\hat{a}_2
		+ \hbar \omega_{m} \hat{b}^{\dagger}\hat{b} \\
		&- \hbar g_{0} \hat{a}_2^{\dagger}\hat{a}_2 (\hat{b} + \hat{b}^{\dagger})
		- \hbar J \left( \hat{a}_1^{\dagger}\hat{a}_2 + \hat{a}_2^{\dagger}\hat{a}_1 \right)
		\end{split}
	\end{equation}
	The first three terms give the evolution of the subsystems when they are isolated, the fourth term creates the interaction between the optical mode in the second ring and the HBAR mechanical mode, and the last term arises from the coupling between the optical modes of the two microrings. 
	The Hamiltonian is then transformed following the procedure described in the Appendix (\ref{appendix:Hamiltonian}) and gives Langevin equations of the internal modes of the converter as
	\footnote{$\Delta_1 = \omega_L - \omega_1$ and $\Delta_2 = \omega_L - \omega_2$ are the detuning of the pump laser to the first and second ring cavities as defined in section (\ref{section:RotatingFrame}). In all generality they do not need to have the same value, even though it was assumed to be the case in section (\ref{section:DeviceImplementation}).}
	\begin{widetext}
		\begin{subnumcases}{}
			&$\frac{d}{dt} \delta \hat{a}_1 =
			i \Delta_1 \delta \hat{a}_1 
			- \frac{\kappa_1}{2} \delta \hat{a}_1
			+ i J \delta \hat{a}_2
			+ \sqrt{\kappa_{0, 1}} \hat{f}_{o, 1}
			+ \sqrt{\kappa_{\text{ex}, 1}} \delta\hat{a}_{\text{in}}$ \\
			&$\frac{d}{dt} \delta \hat{a}_2 =\ 
			i \Delta_2 \delta \hat{a}_2 
			- \frac{\kappa_2}{2} \delta \hat{a}_2
			+ i J \delta \hat{a}_1
			+ i g_{0} \bar{a}_2 (\hat{b} + \hat{b}^{\dagger})
			+ \sqrt{\kappa_{0, 2}} \hat{f}_{o, 2}$ \\
			&$\frac{d}{dt} \hat{b} = 
			-i \omega_{m} \hat{b}  
			- \frac{\gamma_m}{2} \hat{b}
			+ i g_{0} \left( \bar{a}_2^* \delta\hat{a}_2 + \bar{a}_2 \delta\hat{a}_2^{\dagger} \right) 
			+ \sqrt{\gamma_{0}} \hat{f}_{m}
			+ \sqrt{\gamma_{\text{ex}}} \hat{c}_{\text{in}}$
		\end{subnumcases}
	\end{widetext}
	
	First, we solve these linear differential equations in the frequency domain while neglecting the counter-rotating terms. The solution is then more easily understood in terms of transfer functions between the input and output modes.
	Noticing that a few groups of terms appear recurrently in the transfer functions, it is convenient to define the susceptibilities of the optical modes and the mechanical mode
	\begin{align}
		\begin{cases}
			\chi_{o, 1}[s] &= \frac{1}{s - i\Delta_1 + \frac{\kappa_1}{2}} \\
			\chi_{o, 2}[s] &= \frac{1}{s - i\Delta_2 + \frac{\kappa_2}{2}} \\
			\chi_{m}[s] &= \frac{1}{s + i\omega_{m} + \frac{\gamma_m}{2}}
		\end{cases}
	\end{align}
	as well as frequency dependent cooperativities for the optomechanical coupling and for the coupling between the two rings 
	\begin{align}
		\begin{cases}
			C_{\text{OM}}[s] &= \chi_{o, 2}[s] \chi_{m}[s] g_{\text{OM}}^2[s] \\
			C_{\text{OO}}[s] &= \chi_{o, 1}[s] \chi_{o, 2}[s] J^2
		\end{cases}
	\end{align}
	where $g_{\text{OM}}[s] = * g_0\bar{a}_2[s]$\footnote{$g_{\text{OM}}^2[s]$ will thus imply $* g_0^2 \left( \bar{a}_2[s] * \bar{a}_2[s] \right)$.}.
	This contrasts with the usual definition in the optomechanics community where the cooperativity is a constant quantity, but has the advantage of taking into account the frequency dependent behaviour of the total system, and showing that these terms can be understood as coming from the feedback loops between the different modes.
	By further defining frequency dependent extraction efficiencies as
	\begin{align}
		\begin{cases}
			\eta_{\text{opt}}[s] &= \frac{\kappa_{\text{ex}}}{2} \chi_{o, 1}[s] \\
			\eta_{\text{MW}}[s] &= \frac{\gamma_{\text{ex}}}{2} \chi_m[s] 
		\end{cases}
	\end{align}
	and choosing to place the pump laser on the symmetric supermodes of the coupled rings,
	the conversion efficiency from the microwave input to the optical output now takes the simple form
	\begin{equation}
		|G_{ \hat{a}_{\text{out}} \hat{c}_{\text{in}} }[s]|^2 \approx
		\eta_{\text{MW}}[s] \eta_{\text{opt}}[s]
		\frac{4 | C_{\text{OM}}[s] C_{\text{OO}}[s] |}{|C_{\text{OM}}[s] + C_{\text{OO}}[s] + 1|^{2}}
		\label{eq:RWA_efficiency}
	\end{equation}
	The first two factors are the extraction efficiencies, quantifying how well a microwave photon or an optical photon can enter or exit the transducer. These factors are determined by the geometry of the device and are fixed after fabrication. The factor containing the cooperativities is the internal conversion efficiency of the transducer. The internal efficiency increases with the pump power until reaching a maximal value of $100\%$ before slowly decreasing.	
	Fig. \ref{fig:ConversionEfficiency}\textcolor{NavyBlue}{b}) and Fig. \ref{fig:ConversionEfficiency_sweeps}\textcolor{NavyBlue}{c}) show that, for realistic values of the parameters given in the table (\ref{tab:CurrentParameters}), the conversion efficiency is limited by the extraction efficiencies and not the internal conversion process. Using this new piezooptomechanical scheme conversion efficiencies almost reaching $50\%$ should already be within reach of today's technological capabilities.
	
	We note that the form of the formula eq.(\ref{eq:RWA_efficiency}) for the conversion efficiency, which is similar to the results obtained for other converters using three internal modes \cite{wu2020microwave, arnold2020converting}, is slightly different from the one reported in literature for transducers relying on only one optical mode and one microwave mode.
	The additional terms can be understood as arising from the feedback loops created by the bidirectional couplings between the modes. This also implies that this linearized system of equations can be solved from the diagram (\ref{fig:SignalFlowChart}) by using Mason's gain formula \cite{mason1953feedback, mason1956feedback, pozar2011microwave}, similarly to the method introduced by L. Ranzani \textit{et al.} to study nonreciprocity \cite{ranzani2015graph}.
	The complete conversion efficiency accounting for the counter-rotating terms is the square modulus of the transfer function $G_{ \hat{a}_{\text{out}} \hat{c}_{\text{in}} }[s] = \frac{N}{D}$ between the microwave input $\hat{c}_{\text{in}}$ and the optical output $\hat{a}_{\text{out}}$. The numerator of this transfer function is	
	\begin{equation}
		\begin{split}
			N =\ &\sqrt{\gamma_{\text{ex}}} \sqrt{\kappa_{\text{ex}}} \chi_{o, 1}[s] \chi_{o, 2}[s] \chi_{m}[s] J g_{\text{OM}} \\
			&\left( 1 + \chi_{o, 1}^*[s^*] \chi_{o, 2}^*[s^*] J^{2} \right)
		\end{split}
	\label{eq:ConversionNumerator}
	\end{equation}
	and its denominator is
	\begin{align}
		D =\ & 1 
		+ |g_{\text{OM}}|^2 \left( \chi_{o, 2}[s] - \chi_{o, 2}^*[s^*] \right) \left( \chi_m[s] - \chi_m^*[s^*] \right) \nonumber \\
		&+ J^2 \left( \chi_{o, 1}[s] \chi_{o, 2}[s] + \chi_{o, 1}^*[s^*] \chi_{o, 2}^*[s^*] \right) \nonumber \\ 
		&+ 2 |g_{\text{OM}}|^4 \left( \chi_{o, 2}[s] \chi_m[s] \chi_{o, 2}^*[s^*] \chi_m^*[s^*] \right) \nonumber \\
		&+ J^4 \chi_{o, 1}[s] \chi_{o, 2}[s] \chi_{o, 1}^*[s^*] \chi_{o, 2}^*[s^*] \nonumber \\
		&- |g_{\text{OM}}|^2 J^2 \chi_{o, 2}[s] \chi_{o, 2}^*[s^*] \left( \chi_{o, 1}[s] - \chi_{o, 1}^*[s^*] \right) \nonumber \\
		& \ \left( \chi_m[s] - \chi_m^*[s^*] \right)
		\label{eq:ConversionDenominator}
	\end{align}
	This formula is obtained from Fig. \ref{fig:SignalFlowChart}\textcolor{NavyBlue}{a}) by applying Mason's gain rule for signal flow graphs \cite{mason1953feedback}.
	The details of the derivation are reported in the appendix (\ref{section:TotalConversionEfficiency}).
	The validity of this formula does not depend on the detuning of the laser to the optical modes and can be used to analyze the behaviour of the transducer in different regimes of operation, e.g. for amplification by setting the pump laser on the asymmetric supermode.
	The conversion efficiency (\ref{eq:RWA_efficiency}) can be retrieved by neglecting the terms with the conjugate susceptibilities, which is equivalent to applying the rotating wave approximation.
		
	\begin{table}
		\renewcommand{\arraystretch}{1.35} 
		\caption{Values of the parameters used to compute the conversion efficiencies.}
		\begin{tabular}{lll}
			\toprule
			Optical frequency &       $\frac{\omega_c}{2\pi}$ &               \SI{193}{\tera\hertz} \\
			Optical coupling rate &    $\frac{\kappa_{\text{ex}}}{2\pi}$ &               \SI{125}{\mega\hertz} \\
			Intrinsic optical linewidth &       $\frac{\kappa_0}{2\pi}$ &              \SI{25}{\mega\hertz} \\
			Intrinsic optical quality factor &                         $Q_o$ &   $7.5 \times 10^{6}$ \\
			Single photon optomechanical coupling rate &      $\frac{g_{0, \text{OM}}}{2\pi}$ &                \SI{400}{\hertz} \\
			Input laser power &                      $P_{\text{in}}$ &                \SI{100}{\milli\watt} \\
			Number of optical photons &               $|\bar{a}_2|^2$ &   $\mathcal{O}(10^8)$ \\
			Optomechanical coupling rate &         $\frac{g_{\text{OM}}}{2\pi}$ &                \SI{20}{\mega\hertz} \\
			Detuning to the symmetric supermode &       $\frac{\Delta_{\text{S}}}{2\pi}$ &                 \SI{0}{\mega\hertz} \\
			Detuning to the asymmetric supermode &       $\frac{\Delta_{\text{A}}}{2\pi}$ &             \SI{3.285}{\giga\hertz} \\
			Mechanical frequency &       $\frac{\omega_m}{2\pi}$ &             \SI{3.285}{\giga\hertz} \\
			Intrinsic mechanical linewidth &       $\frac{\gamma_0}{2\pi}$ &               \SI{2.6}{\mega\hertz} \\
			Mechanical quality factor &                         $Q_m$ &                   600 \\
			Electromechanical coupling factor   &                   $k_\text{eff}^2$ &  $4.3 \times 10^{-3}$ \\
			Electromechanical coupling rate &         $\frac{g_{\text{EM}}}{2\pi}$ &               \SI{100}{\mega\hertz} \\
			Mechanical coupling rate &  $\frac{|\gamma_{\text{ex}}|}{2\pi}$ &               \SI{2.9}{\mega\hertz} \\
			Total mechanical linewidth &   $\frac{\gamma_\text{total}}{2\pi}$ &               \SI{5.3}{\mega\hertz} \\
			\bottomrule
		\end{tabular}
		\label{tab:CurrentParameters}
	\end{table}

\section{Device implementation}
\label{section:DeviceImplementation}

	Finally, we give an estimation of realistic values of the parameters to assess the feasibility of this approach.
	The values used in this section were obtained from finite elements simulations, and confirmed with preliminary experimental data on devices similar to the one proposed here for transduction.
	Fig. \ref{fig:ConversionEfficiency_sweeps}\textcolor{NavyBlue}{c}) shows the conversion efficiency with the input laser power for different coupling rate of the bus to the ring cavities. It shows that the maximal conversion efficiency achievable is increasing with the bus-ring coupling rate, but the required input power increases similarly. The requirement of not exceeding the cooling power of the fridge for cryogenic operations thus sets the limit to the achievable optical extraction efficiency.
	Fig. \ref{fig:ConversionEfficiency}\textcolor{NavyBlue}{b}) shows the cross-section of Fig. \ref{fig:ConversionEfficiency_sweeps}\textcolor{NavyBlue}{c}) for a bus-ring coupling rate $\frac{\kappa_{\text{ex}}}{2\pi} = \SI{125}{\mega\hertz}$. The corresponding parameter values can easily be reached on devices fabricated with current CMOS-compatible technologies and achieves up to $40\%$ conversion efficiency between the microwave and optical domains with a reasonable input laser power on the symmetric supermode of \SI{100}{\milli\watt}.
	The table \ref{tab:ComparativeTable} shows that the approach proposed here competes with the state-of-the-art of microwave-optical transducers. 
	While the theoretical results presented here were obtained in the assumption of a continuous wave pump laser, the same conclusions would applied in the case of a pulsed pump, at least for a repetition rate up to a few \si{\mega\hertz}.
	While direct quantum coherent transduction requires slight technological improvements, the proposed scheme provides high conversion efficiencies that are compatible with protocols guarranteeing quantum coherent conversion \cite{Lau2019, wu_deterministic_2021}.
	
	\begin{table*}[htb]
		\caption{Comparison of state-of-the-art microwave-to-optical transducers.}
		\begin{tabular}{lllllllll}
			\toprule
			Reference &         Device type &            Input power &      Frequency &      On-chip efficiency &  Normalized efficiency &             Temperature & Laser scheme & Material \\
			\midrule
			\cite{fan2018superconducting} &                  EO &    \SI{6}{\milli\watt} &   \SI{8}{\giga\hertz} &   $2.05 \times 10^{-2}$ &    \SI{3.4}{\per\watt} &       \SI{1.7}{\kelvin} &           CW &           \ch{AlN} \\
			\cite{fu2020ground} &                  EO &   \SI{72}{\micro\watt} &   \SI{6}{\giga\hertz} &    $2.4 \times 10^{-5}$ &    \SI{0.3}{\per\watt} &  \SI{40}{\milli\kelvin} &       Pulsed &           \ch{AlN} \\
			\cite{xu2020bidirectional} &                  EO &   \SI{20}{\milli\watt} &   \SI{8}{\giga\hertz} &   $1.02 \times 10^{-2}$ &    \SI{0.5}{\per\watt} &       \SI{1.9}{\kelvin} &       Pulsed &        \ch{LiNbO3} \\
			\cite{wu2020microwave} & PiezoOM &   \SI{107}{\nano\watt} &   \SI{3}{\giga\hertz} & $<< 2.5 \times 10^{-5}$ &  \SI{233.6}{\per\watt} &  \SI{20}{\milli\kelvin} &       Pulsed &          \ch{GaAs} \\
			\cite{andrews2014bidirectional} &                  OM &                     -  &   \SI{7}{\giga\hertz} &    $8.6 \times 10^{-2}$ &                      - &         \SI{4}{\kelvin} &           CW &         Free space \\
			\cite{higginbotham2018harnessing} &                  OM &                     -  &   \SI{1}{\mega\hertz} &    $4.7 \times 10^{-1}$ &                      - &  \SI{35}{\milli\kelvin} &           CW &         Free space \\
			\cite{rueda2016efficient} &                  EO & \SI{0.42}{\milli\watt} &   \SI{9}{\giga\hertz} &   $1.09 \times 10^{-3}$ &    \SI{2.6}{\per\watt} &                      RT &           CW &        \ch{LiNbO3} \\
			\cite{mirhosseini2020quantum} & PiezoOM &   \SI{20}{\micro\watt} &   \SI{5}{\giga\hertz} &    $8.8 \times 10^{-6}$ &    \SI{0.4}{\per\watt} &  \SI{15}{\milli\kelvin} &       Pulsed &            \ch{Si} \\
			\cite{arnold2020converting} &                  OM &   \SI{625}{\pico\watt} &  \SI{10}{\mega\hertz} &    $1.2 \times 10^{-2}$ &  \SI{1.9e8}{\per\watt} &  \SI{50}{\milli\kelvin} &           CW &            \ch{Si} \\
			\cite{hease2020bidirectional} &                  EO & \SI{1.48}{\milli\watt} &   \SI{9}{\giga\hertz} &   $3.16 \times 10^{-4}$ &    \SI{0.2}{\per\watt} & \SI{320}{\milli\kelvin} &           CW &        \ch{LiNbO3} \\
			\cite{holzgrafe2020cavity} &                  EO &                     -  &   \SI{3}{\giga\hertz} &    $2.7 \times 10^{-5}$ &                      - &         \SI{1}{\kelvin} &           CW &        \ch{LiNbO3} \\
			\cite{rueda2016efficient} & PiezoOM &  \SI{3.3}{\micro\watt} &   \SI{2}{\giga\hertz} &      $1 \times 10^{-5}$ &    \SI{3.0}{\per\watt} &                      RT &           CW &        \ch{LiNbO3} \\
			\cite{yamazaki2020radio} & PiezoOM &  \SI{372}{\micro\watt} & \SI{300}{\mega\hertz} &    $1.2 \times 10^{-6}$ & \SI{3.0e-3}{\per\watt} &                      RT &           CW &        \ch{LiNbO3} \\
			\cite{stockill2021ultra} & PiezoOM &                      - &   \SI{3}{\giga\hertz} &    $6.8 \times 10^{-8}$ &                      - &  \SI{20}{\milli\kelvin} &       Pulsed &           \ch{GaP} \\
			This work & PiezoOM &  \SI{100}{\milli\watt} &   \SI{3}{\giga\hertz} &    $4.0 \times 10^{-1}$ &    \SI{4.0}{\per\watt} &         \SI{3}{\kelvin} &       Pulsed &         \ch{Si3N4} \\
			\bottomrule
		\end{tabular}
		\label{tab:ComparativeTable}
	\end{table*}
	
	The efficiency is not limited by the internal conversion process, and can thus be improved by overcoupling the optical and mechanical modes to their input baths.
	This can be done by optimizing the position of the bus waveguide relatively to the coupled rings while achieving lower optical loss, or optimizing the geometry of the HBAR resonator to increase the electromechanical coupling factor. Using another piezoelectric material with a better electromechanical coupling factor $K^2$, such as \ch{Sc} doped \ch{AlN}, is another viable option to reach a higher $k_{\text{eff}}^2$, and thus improving the microwave-to-mechanics extraction efficiency.

\section{Conclusion}

	With this paper we propose a new type of transducer to convert travelling photons between the microwave and optical domains, harnessing the synergy between photonic integrated circuit and MEMS technologies that are fully compatible with high-yield wafer-scale manufacturing.
	The results of the theoretical analysis show that conversion efficiencies above $50\%$, the threshold required to apply quantum error correction techniques over the microwave-optical link, will soon be within reach  \cite{fu2020ground}.
	We furthermore presented a general method to solve the dynamics of a multipartite quantum system in order to assess its performance as mode converter.
	While significant optical power is still required to reach the maximal internal conversion efficiency, improvements in the electromechanical coupling can easily be achieved by using materials with stronger piezoelectric behaviour, such as \ch{ScAlN} or even \ch{LiNbO3}. Nevertheless, this transducer could readily be used even with power much lower than what is required for maximizing the conversion efficiency, for example, for generating entangled pairs of microwave and optical photons, or for amplifying weak microwave signals that would be read out with a photodetector.

\begin{acknowledgments}
	The authors are thankful to Liu Qiu for fruitful discussions on the theory of optomechanics and Anat Siddharth for insights about experimental aspects of the measurements of HBAR devices.
	This material is based upon work supported by the Air Force Office of Scientific Research under number FA9550-21-1-0047 (Quantum Accelerator), as well as NSF QISE-Net under grant DMR 17-47426.
	This work was further supported by funding from the European Union H2020 research and innovation programme under grant agreement No. 732894 (FET-Proactive HOT)), and the European Research Council (ERC) under grant agreement No. 835329 (ExCOM-Cceo).	
\end{acknowledgments}


\appendix

\section{Lagrangian formulation of the piezoelectric fields}
\label{section:Lagrangian}

	\subsection{Lagrangians of the uncoupled fields}

		The derivation of the electromechanical coupling rate from the piezoelectric interaction requires to quantize the mechanical and electromagnetic modes inside the piezoelectric material. This can be done after obtaining the proper Hamiltonian densities.
		To derive the Hamiltonian of the fields it is useful to first go through their Lagrangian formulation.
		
		The Lagrangian enables to describe the evolution of a system for a given set of generalized coordinates and their derivatives $\{ \phi_i, \phi_{i,j} \}$
		\begin{equation}
			L(\phi_i, \phi_{i,j}) = T - V = \iiint \mathcal{L}(\phi_i, \phi_{i,j})\ d\mathcal{V}
		\end{equation}
		thanks to Euler-Lagrange's equations
		\begin{equation}
			\frac{\partial \mathcal{L}(\phi_i, \phi_{i,j})}{\partial \phi_i} = \sum_j \frac{\partial}{\partial x_j} \left( \frac{\partial \mathcal{L}(\phi_i, \phi_{i,j})}{\partial \phi_{i,j}} \right)
			\label{eq:EulerLagrange}
		\end{equation}
		Lagrangian density of the elastic waves is
		\begin{equation}
			\mathcal{L}_{\text{e.w.}} = \sum_i \frac{1}{2} \rho \dot{u}_i \dot{u}_i - \sum_{ijkl} \frac{1}{2} c_{ijkl} u_{i,j} u_{k,l}
		\end{equation}
		for which the corresponding Hamiltonian density
		\begin{equation}
			\mathcal{H}_{\text{e.w.}} = \sum_{i} \frac{1}{2} \rho \dot{u}_i \dot{u}_i + \sum_{ijkl} \frac{1}{2} c_{ijkl} u_{i,j} u_{k,l}
		\end{equation}
		is derived with Legendre's transformation $\mathcal{H} = \sum_i \dot{\phi}_i \frac{\partial \mathcal{L}}{\partial \dot{\phi}_i }  - \mathcal{L}$ \cite{laude2015lagrangian}.
		By the same principle, the Lagrangian density of the electromagnetic waves in presence of external charges and currents
		\begin{equation}
		\mathcal{L}_{\text{EM}} = - \rho_e \phi_e + \mathbf{j}\cdot\mathbf{A} + \frac{\epsilon_0 \mathbf{E}^2}{2} - \frac{\mathbf{B}^2}{2\mu_0}
		\end{equation}
		leads to the Hamiltonian density \cite{laude2015lagrangian}
		\begin{equation}
		\mathcal{H}_{\text{EM}} = \frac{1}{2} \left( \mathbf{E}\cdot\mathbf{D} + \mathbf{H}\cdot\mathbf{B} \right)
		\end{equation}
	
	\subsection{Piezoelectric coupling}

		A Lagrangian describing a system is only legitimate if it reproduces the known equations of motion.
		In the case of piezoelectric materials, these equations of motion are the conservation of linear momentum
		\begin{equation}
			\rho \ddot{u}_i = \sum_j T_{ij,j}
		\end{equation}
		and the conservation of charges in the medium
		\begin{equation}
			\sum_{i} D_{i,i} = 0
			\label{eq:ChargeConservation}
		\end{equation}
		with the stress and the electric displacement given by eq.(\ref{eq:PiezoStress}) and eq.(\ref{eq:PiezoDisplacement}), together with Maxwell's equations.
		
		The conservation of momentum translates to
		\begin{equation}
			\rho \ddot{u}_i = \sum_{kl} c_{ijkl} S_{kl, j} - \sum_{k} e_{kij} E_{k, j}
			\label{eq:MomentumConservation}
		\end{equation}
		A Lagrangian density can then be constructed knowing that (\ref{eq:MomentumConservation}) has to arise from (\ref{eq:EulerLagrange}).
		We can see by direct identification that
		\begin{equation}
			\frac{\partial \mathcal{L}}{\partial \dot{u}_i} = \rho \dot{u}_i
		\end{equation}
		will lead to the mechanical kinetic energy term $\frac{1}{2}\rho \dot{u}_i^2$ up to some constant, and that
		\begin{equation}
			\frac{\partial}{\partial x_j} \left( \frac{\partial \mathcal{L}}{\partial u_{i,j}} \right) 
			= \frac{\partial}{\partial x_j} \left( c_{ijkl} u_{k,l} - e_{kij}E_k \right)
		\end{equation}
		will lead to potential terms $-\frac{1}{2} c_{ijkl} u_{i,j}u_{k,l} + e_{kij} E_k u_{i,j}$, up to some constants.
		This form of the elastic potential energy strongly suggests that the piezoelectric interaction is produced by adding a term $e_{kij} E_k S_{ij}$ in the Lagrangian density. It is then required to check if a Lagrangian density constructed with this term will still obey Maxwell's equations while reproducing the constitutive relations for linear piezoelectric matierals.
		
		We follow the reasoning exposed in \cite{hillery_introduction_2009} in order to check the validity of the new Lagrangian density
		\begin{align}
			\mathcal{L} &= 
			\sum_{i} \frac{1}{2} \rho \dot{u}_i \dot{u}_i
			- \sum_{ijkl} \frac{1}{2} c_{ijkl} u_{i,j} u_{k,l} \nonumber \\
			& + \sum_{ij} \frac{1}{2} \epsilon_{ij} E_i E_j
			+ \sum_{ijk} e_{kij} E_k S_{ij}
			\label{eq:PiezoLagrangian}
		\end{align}
		The Lagrangian density for electromagnetic fields in a dielectric is known to be $\mathcal{L}_{\mathrm{EM}} = \sum_i \frac{1}{2} E_i \epsilon_{ij} E_j - \sum_{ij} \frac{1}{2} B_{i} \frac{1}{\mu_{ij}} B_j$.
		Two of Maxwell's equations, $\nabla \cdot \mathbf{B} = 0$ and $\nabla \times \mathbf{E} = - \frac{\partial \mathbf{B}}{\partial t}$, can be automatically satisfied by choosing to work with the scalar potential $A_0$ and the vector potential $\mathbf{A}$, related to the electric and magnetic fields through $\mathbf{E} = - \frac{\partial \mathbf{A}}{\partial t} - \nabla A_0$ and $\mathbf{B} = \nabla \times \mathbf{A}$.
		The Euler-Lagrange equations (\ref{eq:EulerLagrange}) for $i = 1, 2, 3$ are related to Amp\`ere's law $\nabla \times \mathbf{B} = \mu \left( \mathbf{J} + \frac{\partial \mathbf{D}}{\partial t} \right)$. The development is not shown here as it has no important consequence for the quantization of the piezoelectric interaction, but the reasoning is similar to what is done for the last component.
		The Euler-Lagrange equation (\ref{eq:EulerLagrange}) for the scalar component of the potential $i = 0$ 
		\begin{equation}
			\frac{\partial \mathcal{L}}{\partial A_0} = \frac{d}{dt} \frac{\partial \mathcal{L}}{\partial \frac{\partial A_0}{\partial t}} + \sum_{j=1}^3 \frac{\partial}{\partial x_j} \frac{\partial \mathcal{L}}{\partial \frac{\partial A_0}{\partial x_j}}
			\label{eq:ScalarPotential_EOM}
		\end{equation}
		is the one relevant for verifying (\ref{eq:ChargeConservation}) as it leads to Gauss'law $\nabla \cdot \mathbf{D} = \rho_{\mathrm{free}}$, and is therefore linked to charge conservation.
		Focusing on the parts of the Lagrangian (\ref{eq:PiezoLagrangian}) containing the electric field
		\begin{align*}
			\mathcal{L}_{\mathrm{electric}^{\mathrm{piezo}}} &= \sum_{ij} \frac{1}{2} E_i \epsilon_{ij} E_j - \sum_{ijk} e_{kij} E_{k} S_{ij} \\
			&= \sum_{ij} \frac{1}{2} \epsilon_{ij} (A_{0,i}^2 + A_{i, 0}^2 + 2 A_{0, i}A_{i, 0}) \\
			&\ + \sum_{ijk} e_{kij} \left( -A_{k,0} - A_{0,k}  \right) S_{ij}
		\end{align*}
		we apply (\ref{eq:ScalarPotential_EOM}) to find
		\begin{align*}
			& \sum_{k=1}^3 \frac{\partial}{\partial x_i} \left( \epsilon_{kl}A_{0,l} + \epsilon_{kl} A_{l,0} - e_{kij}S_{ij} \right) = 0 \\
			\Leftrightarrow & \sum_{k=1}^3 \frac{\partial}{\partial x_i} \left( \epsilon_{kl} E_l + e_{kij}S_{ij} \right) = 0
		\end{align*}
		which is equivalent to charge conservation with a displacement electric field given by the constitutive equation eq.(\ref{eq:PiezoDisplacement}) for piezoelectric materials.
		(\ref{eq:PiezoLagrangian}) is thus a valid Lagrangian density for starting the piezoelectric interaction quantization as it reproduces the laws of continuum mechanics and Maxwell's equations, while adding electromechanical coupling through piezoelectricity.
		
		Finally, the corresponding Hamiltonian density is obtained taking a Legendre transform on the coordinates $\{ u_i, E_i \}$, leading to
		\begin{align}
			\mathcal{H} &=
			\sum_{i} \frac{1}{2} \rho \dot{u}_i \dot{u}_i
			+ \sum_{ijkl} \frac{1}{2} c_{ijkl} u_{i,j} u_{k,l} \nonumber \\
			&- \sum_{ij} \frac{1}{2} \epsilon_{ij} E_i E_j
			- \sum_{ijk} e_{kij} E_k S_{ij}
		\end{align}
		Interestingly, this Hamiltonian exactly coincides with the electric enthalpy $H_E(S_{ij}, E_i) = U(S_{ij}, D_i) - \mathbf{E}\cdot\mathbf{D}$ \cite{ieee1984american}.
		This is due to the fact that the internal energy needs to be expressed in terms of extensive variables, here the components of $\mathbf{D}$, while the operation of piezoelectric devices is done by controlling the electric field components through the voltage, which are intensive variables. A Legendre transform hence has to be applied on the variables $\{ D_i \}$ to change the electrical coordinates to $\{ E_i \}$, a more natural choice for studying piezoelectric coupling.

\section{Transformation of the Hamiltonian}
\label{appendix:Hamiltonian}
	
	\subsection{Unitary transformation}
	
	In this appendix, we justify the formula used to apply a unitary transformation to the Hamiltonian.
	
	Starting from Schrodinger equation
	\begin{equation}
	i \hbar \frac{\partial}{\partial t} \ket{\psi} = \hat{H} \ket{\psi}
	\end{equation}
	and the new state $\ket{\tilde{\psi}} = \hat{U} \ket{\psi}$ obtained after applying the unitary transformation $\hat{U}$,
	we obtain
	\begin{equation}
	\begin{split}
	& i \hbar \frac{\partial}{\partial t} \left( \hat{U}^{\dagger} \ket{\tilde{\psi}} \right) = \hat{H} \hat{U}^{\dagger} \ket{\tilde{\psi}} \\
	\Leftrightarrow \quad & i \hbar \frac{\partial}{\partial t} \left( \hat{U}^{\dagger} \right) \ket{\tilde{\psi}}
	+
	i \hbar \hat{U}^{\dagger} \frac{\partial}{\partial t} \left(  \ket{\tilde{\psi}} \right) 
	= \hat{H} \hat{U}^{\dagger} \ket{\tilde{\psi}} \\
	\Leftrightarrow \quad & i \hbar \hat{U}^{\dagger} \frac{\partial}{\partial t} \left(  \ket{\tilde{\psi}} \right) 
	= 
	\hat{H} \hat{U}^{\dagger} \ket{\tilde{\psi}}
	- i \hbar \frac{\partial}{\partial t} \left( \hat{U}^{\dagger} \right) \ket{\tilde{\psi}} \\
	\Leftrightarrow \quad & i \hbar \frac{\partial}{\partial t} \left(  \ket{\tilde{\psi}} \right) 
	= 
	\hat{U} \hat{H} \hat{U}^{\dagger} \ket{\tilde{\psi}}
	- i \hbar \hat{U} \frac{\partial}{\partial t} \left( \hat{U}^{\dagger} \right) \ket{\tilde{\psi}}
	\end{split}
	\end{equation}
	which justifies the form of the Hamiltonian in the new frame
	\begin{align}
	\hat{H}^{\text{new}} = \hat{U} \hat{H}^{\text{old}} \hat{U}^{\dagger} - i \hbar \hat{U} \frac{\partial \hat{U}^{\dagger}}{\partial t}
	\end{align} \\
	
	\subsection{Rotating frame}
	\label{section:RotatingFrame}
	
	The form of the Hamiltonian in the frame rotating at the frequency of the pump laser $\omega_L$ is computed using the relation derived in the previous section.
	
	In this case the transformation is given by $\hat{U} = e^{i \hat{a}^{\dagger}\hat{a} \omega_L t}$ so that $\ket{\psi^{\text{old}}} = e^{- i \hat{a}^{\dagger}\hat{a} \omega_L t} \ket{\psi^{\text{new}}}$, i.e. the optical wavefunctions in the new frame correspond to slowly varying envelopes.
	Using the fact that
	\begin{equation}
	\hat{U} (\hbar \omega_c \hat{a}^{\dagger}\hat{a}) \hat{U}^{\dagger} 
	= \hbar \omega_c \hat{a}^{\dagger}\hat{a}
	\end{equation}
	and
	\begin{equation}
	\frac{\partial \hat{U}^{\dagger}}{\partial t}
	= - i \omega_L \hat{a}^{\dagger} \hat{a}\ \hat{U}^{\dagger}
	\end{equation}
	the effect of applying $\hat{U} = e^{i \omega_L (\hat{a}_1^{\dagger}\hat{a}_1 + \hat{a}_2^{\dagger}\hat{a}_2) t}$ to move in the rotating frame is only to transform the terms from the isolated optical subsystems into
	\begin{equation}
	\hat{H}^{\text{new}} = 
	\hbar \omega_c \hat{a}^{\dagger}\hat{a}
	- \hbar \omega_L \hat{a}^{\dagger}\hat{a}
	= - \hbar \Delta \hat{a}^{\dagger}\hat{a}
	\end{equation}
	where we defined the detuning of the optical mode to the pump laser as $\Delta = \omega_L - \omega_c$.
	
	\subsection{Optomechanical interaction linearization}
	
	The optomechanical interaction Hamiltonian \cite{aspelmeyer2014cavity}
	\begin{equation}
	\hat{H}_{\text{OM}} = 
	- \hbar g_0 \hat{a}^{\dagger}\hat{a} (\hat{b} + \hat{b}^{\dagger})
	\end{equation}
	reveals itself to be problematic when looking at the evolution of annihilation operators because of the triple product, that leaves nonlinear terms in the Langevin equations.
	Conventional solution to this problem used by the community of researchers in the field of optomechanics is to linearize the optical field amplitude around the average value of the field in the cavity $\bar{a} = \expval{\hat{a}}$.
	This approximation consists in separating the mean value and the fluctuations, which take into account for the quantum properties we are interested in with this transduction scheme proposal,
	\begin{equation}
	\hat{a} \approx \bar{a} + \delta\hat{a}
	\end{equation}
	With this expansion of the optical field, the optomechanical Hamiltonian becomes
	\begin{equation}
	\begin{split}
	&\hbar g_0 \hat{a}^{\dagger}\hat{a} (\hat{b} + \hat{b}^{\dagger})
	\approx \\
	&\hbar g_0
	\left[ |\bar{a}|^2 + \delta\hat{a}^{\dagger}\delta\hat{a} + \bar{a} \delta\hat{a}^{\dagger} 
	+ \bar{a}^* \delta\hat{a} \right]
	(\hat{b} + \hat{b}^{\dagger})
	\end{split}
	\end{equation}
	where we can identify three kind of terms:
	\begin{enumerate}
		\item a constant term causing a shift of the displacement origin $\hbar g_0 |\bar{a}|^2 (\hat{b} + \hat{b}^{\dagger})$
		\item a term that will be kept as the linearized optomechanical Hamiltonian $\hbar g_0 \left[ \bar{a} \delta\hat{a}^{\dagger} + \bar{a}^* \delta\hat{a} \right] (\hat{b} + \hat{b}^{\dagger})$
		\item and a term $\hbar g_0 \delta\hat{a}^{\dagger}\delta\hat{a} (\hat{b} + \hat{b}^{\dagger})$ that is neglected for having a contribution at least $|\bar{a}|$ smaller than the two other, which rapidly becomes consequent for reasonable pump power.
	\end{enumerate}
	The first disappears by applying an appropriate shift $\delta x$ of the displacement origin.
	The value of this shift is obtained by evaluating the susceptibility of the mechanical oscillator in the steady state $\chi[0] = \frac{1}{k} = \frac{1}{m_{\text{eff}} \omega_m^2}$ multiplied by the average radiation pressure force $\bar{F}_{\text{rad. press.}} = \hbar G |\bar{a}|^2$, giving the displacement at equilibrium
	\begin{equation}
	\delta x = \frac{\hbar G |\bar{a}|^2}{m_{\text{eff}}\omega_m^2}
	\end{equation}
	for a frequency shift per displacement unit $G = - \frac{\partial \omega_{\text{cav}}}{\partial x} = \frac{g_0}{x_{\text{ZPF}}}$.
	Writing the displacement and the associated momentum in terms of annihilation operators
	\begin{align}
	\begin{cases}
	\hat{x} = \sqrt{\frac{\hbar}{2 m_{\text{eff}} \omega_m}} (\hat{b} + \hat{b}^{\dagger}) \\
	\hat{p}_x = \sqrt{\frac{\hbar m_{\text{eff}} \omega_m}{2}} \frac{1}{i} (\hat{b} - \hat{b}^{\dagger})
	\end{cases}
	\end{align}
	we can then easily construct the appropriate translation operator as \cite{basdevant2002mecanique}
	\begin{equation}
	\hat{T}(\delta x) = e^{\frac{1}{i\hbar} \delta x \hat{p}_x}
	\end{equation}
	Applying this operator to the original Hamiltonian, we find that the first term occurring in the linearization of the optomechanical Hamiltonian vanishes at the cost of shifting the optical resonances frequency, leading to a modified detuning
	\begin{equation}
	\Delta_{\text{new}} = \Delta_{\text{old}} + G \delta\bar{x}
	\end{equation} \\
	
	\subsection{Intracavity fields}
	\label{subsection:CavityFields}
	
	\begin{figure*}
		\includegraphics[width=0.95\textwidth]{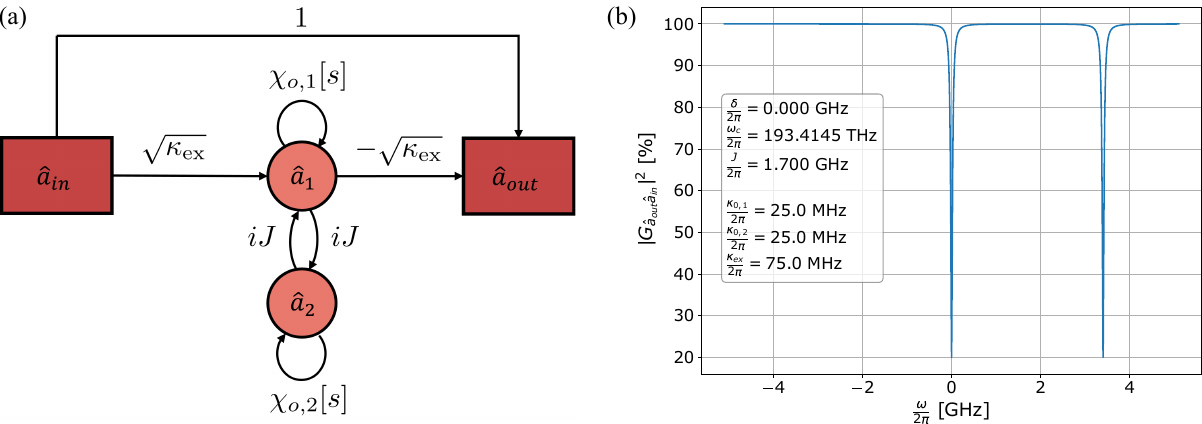}
		\caption{\textbf{Optical transmission of the coupled microrings.}
				(a)
				Signal flow chart describing the behaviour of the two coupled optical modes, without the effect of the mechanics. The transfer functions of the intracavity fields can also be obtained by applying Mason's gain rule to this diagram.
				(b)
				Optical transmission of the coupled rings system. Both rings are set to have the same frequency $\omega_c$ (there is on detuning $\delta$ between the rings resonant frequency). The strong coupling $J$ between them causes modes hybridization, which results in two effective resonance frequency separated by $2 J$.
				The two absorption peaks corresponds to the symmetric (lower frequency) and asymmetric supermodes of the photonic dimer.
				The transmission spectrum is showed in the frame rotating at $\omega_c$.}
		\label{fig:SignalFlowChart_Optical}
		\label{fig:OpticalTransmission}
	\end{figure*}
	
	In this section we explain how to obtain the intracavity photon numbers for the two rings of the photonic dimer, terms that arise in the optomechanical coupling rate.
	
	We consider a pair of coupled optical modes from which only the first one can be probed via a bus waveguide. This situation is illustrated with the signal flow graph on Fig. \ref{fig:SignalFlowChart_Optical}.
	Linearizing the annihilation operators of the optical modes as described in the previous section, we can separate the evolution of the fluctuations from the evolution of the average cavity field
	\begin{align}
		\frac{d}{dt} \hat{a}_1 = 
		\frac{d}{dt} \bar{a}_1
		+ \frac{d}{dt} \delta\hat{a}_1 \\
		\frac{d}{dt} \hat{a}_2 = 
		\frac{d}{dt} \bar{a}_2
		+ \frac{d}{dt} \delta\hat{a}_2
	\end{align}
	We group the terms with $\delta\hat{ a }_1$ and $\delta\hat{ a }_2$ to obtain the Langevin equations for the fluctuations operators
	\begin{align}
		\frac{d}{dt} \delta \hat{a}_1 =\ 
		&i \Delta_1 \delta \hat{a}_1 
		- \frac{\kappa_1}{2} \delta \hat{a}_1
		+ i J \delta \hat{a}_2 \nonumber \\
		&+ \sqrt{\kappa_{0, 1}} \hat{f}_{o, 1}
		+ \sqrt{\kappa_{\text{ex}, 1}} \delta\hat{a}_{\text{in}} \\
		\frac{d}{dt} \delta \hat{a}_2 =\ 
		&i \Delta_2 \delta \hat{a}_2 
		- \frac{\kappa_2}{2} \delta \hat{a}_2
		+ i J \delta \hat{a}_1 \nonumber \\
		&+ i g_{0} \bar{a}_2 (\hat{b} + \hat{b}^{\dagger})
		+ \sqrt{\kappa_{0, 2}} \hat{f}_{o, 2}
	\end{align}
	while the remaining terms give the equation of motion of the average intracavity fields
	\begin{align}
		\frac{d}{dt} \bar{a}_1 &= -i \omega_1 \bar{a}_1 
		- \frac{\kappa_1}{2} \bar{a}_1
		+ i J \bar{a}_2
		+ \sqrt{\kappa_{\text{ex}, 1}} \bar{a}_{\text{in}} \\
		\frac{d}{dt} \bar{a}_2 &= 
		-i \omega_2 \bar{a}_2 
		- \frac{\kappa_2}{2} \bar{a}_2
		+ i J \bar{a}_1 
	\end{align}
	Taking Laplace's transform, we then obtain the values of the intracavity fields in the frequency domain
	\begin{align}
		\begin{cases}
			\bar{a}_1[s] &=
			\sqrt{\kappa_{\text{ex}}}
			\frac{ \chi_{o, 1}[s] }{ \left(1
				+ J^2 \chi_{o, 1}[s] \chi_{o, 2}[s] \right) } \bar{a}_{\text{in}}[s] \\
			\bar{a}_2[s] &=
			i J \chi_{o, 2}[s] \bar{a}_1[s]
		\end{cases}
	\end{align}
	The output spectrum corresponding to these intracavity fields is shown on Fig. \ref{fig:OpticalTransmission}.

	\subsection{Input laser spectrum}
	\label{subsection:LaserSpectrum}
	
	We assume an input laser with a Lorentzian lane shape, with linewidth $k_L$ and total power $P_{\text{in}}$.
	
	Setting the laser spectrum to be
	\begin{equation}
		\bar{a}_{\text{in}}[-i\omega] = \frac{\frac{\kappa_L}{2}}{-i (\omega-\omega_L) + \frac{\kappa_L}{2}} 
		\sqrt{ \frac{P_{\mathrm{in}}}{\hbar \omega_L} }
		e^{i \phi_L}
	\end{equation}
	it possesses the correct linewidth and gives the correct total number of photons
	\begin{equation}
		|\bar{a}_{\text{in}}[-i\omega_L]|^2
		= \frac{P_{\text{in}}}{\hbar \omega_L}
	\end{equation}
	Taking, for example, a total power of \SI{10}{\milli\watt} and a linewidth of $\frac{\kappa_L}{2\pi} = \SI{10}{\kilo\hertz}$ for the pump laser, which are realistic values, its spectrum is very narrow compared to all the other \si{MHz} order linewidth of the modes in the transducer. This justifies that we assume a Dirac peak for the laser spectrum during the computation of the conversion efficiency.

\section{Transfer functions}
\label{appendix:TransferFunctions}

	In this appendix we give a brief reminder on notions of control theory applied here to solve the linear system of equations describing the transducer.
	The complete set of transfer functions is given applying this state space model, while neglecting the counter-rotating terms
	The derivation of the microwave-optical transfer function with counter-rotating terms is obtained using the signal flow graph \ref{fig:SignalFlowChart} and Mason's gain rule.

	\subsection{State space representation}
	
	The Langevin equation obtained by correspondance rules from Poisson's bracket \cite{basdevant2002mecanique}
	\begin{equation}
		\frac{d}{dt} \hat{a} = \frac{1}{i\hbar} \left[ \hat{a} , \hat{H} \right] + \frac{\partial \hat{a}}{\partial t}
	\end{equation}
	is applied to the Hamiltonian (\ref{eq:Hamiltonian}) to obtain the equation of motion of the internal modes of the transducer.
	Neglecting the counter-rotating terms that are not playing a significant role here, the Langevin equations describing the conversion process of the transducer are	
	\begin{align}
		\frac{d}{dt} \delta \hat{a}_1 &=
		i \Delta_1 \delta \hat{a}_1 
		- \frac{\kappa_1}{2} \delta \hat{a}_1
		+ i J \delta \hat{a}_2 \nonumber \\
		&+ \sqrt{\kappa_{0, 1}} \hat{f}_{o, 1}
		+ \sqrt{\kappa_{\text{ex}, 1}} \delta\hat{a}_{\text{in}} \\
		\frac{d}{dt} \delta \hat{a}_2 &=
		i \Delta_2 \delta \hat{a}_2 
		- \frac{\kappa_2}{2} \delta \hat{a}_2 \nonumber \\
		&+ i J \delta \hat{a}_1
		+ i g_{0} \bar{a}_2 \hat{b}
		+ \sqrt{\kappa_{0, 2}} \hat{f}_{o, 2} \\
		\frac{d}{dt} \hat{b} &=
		-i \omega_{m} \hat{b}  
		- \frac{\gamma_m}{2} \hat{b}
		+ i g_{0} \bar{a}_2^* \delta\hat{a}_2 \nonumber \\
		&+ \sqrt{\gamma_{0}} \hat{f}_{m}
		+ \sqrt{\gamma_{\text{ex}}} \hat{c}_{\text{in}}
	\end{align}
	
	It is convenient to write this linear system of differential equations in a matrix form
	\begin{equation}
		\dot{\matr{x}} = \matr{A} \matr{x} + \matr{B} \matr{u}
	\end{equation} 
	with the state vector
	\begin{equation}
		\matr{x} =
		\begin{pmatrix}
		\delta\hat{a}_1 \\
		\delta\hat{a}_2 \\
		\hat{b}
		\end{pmatrix}
	\end{equation}
	and the input vector 
	\begin{equation}
		\matr{u} =
		\begin{pmatrix}
		\hat{a}_{\text{in}} \\
		\hat{c}_{\text{in}} \\
		\hat{f}_{o, 1} \\
		\hat{f}_{o, 2} \\
		\hat{f}_{m}
		\end{pmatrix}
	\end{equation}
	where the first two terms are the input fields in the device, and the remaining three comes from the noises.
	The system matrix is then given by
	\begin{equation}
		\matr{A} =
		\begin{pmatrix}
		i \Delta_1 - \frac{\kappa_1}{2} & i J & 0\\ \\
		i J & i \Delta_2 - \frac{\kappa_2}{2} & i g_{\text{OM}}\\ \\
		0 & i g_{\text{OM}} & - \frac{\gamma}{2} - i \omega_{m}
		\end{pmatrix}
	\end{equation}
	and the input matrix is
	\begin{equation}
		\matr{B} =
		\left(\begin{matrix}
		\sqrt{\kappa_{\text{ex}}} & 0 & \sqrt{\kappa_{0, 1}} & 0 & 0\\ \\
		0 & 0 & 0 & \sqrt{\kappa_{0, 2}} & 0\\ \\
		0 & \sqrt{\gamma_{\text{ex}}} & 0 & 0 & \sqrt{\gamma_0}\end{matrix}\right)
	\end{equation}
	The input-output relations \cite{caves1982quantum}
	\begin{align}
	\hat{a}_{\text{out}} = \hat{a}_{\text{in}} - \sqrt{\kappa_{\text{ex}}} \hat{a}_1 \\
	\hat{c}_{\text{out}} = -\hat{c}_{\text{in}} + \sqrt{\gamma_{\text{ex}}} \hat{b}
	\end{align}
	then give the output matrix
	\begin{equation}
	\matr{C} = 
	\begin{pmatrix}
	-\sqrt{\kappa_{\text{ex}}} & 0 & 0 \\ \\
	0 & 0 & \sqrt{\gamma_{\text{ex}}}
	\end{pmatrix}
	\end{equation}
	and the feedthrough matrix
	\begin{equation}
	\matr{D} = 
	\begin{pmatrix}
	1 & 0 & 0 & 0 & 0  \\ \\
	0 & -1 & 0 & 0 & 0 
	\end{pmatrix}
	\end{equation}
	for an output vector 
	\begin{equation}
	\matr{y} =
	\begin{pmatrix}
	\hat{a}_{\text{out}} \\
	\hat{c}_{\text{out}}
	\end{pmatrix}
	= \matr{C} \matr{x} + \matr{D} \matr{u}
	\end{equation}
	Taking Laplace's transformation we have in the frequency domain
	\begin{equation}
	\begin{split}
	s \tilde{\matr{x}}[s] - \matr{x}(0) &= \matr{A} \tilde{\matr{x}}[s] + \matr{B} \tilde{\matr{u}}[s] \\
	\Leftrightarrow
	\tilde{\matr{x}}[s] &= (s\matr{I} - \matr{A})^{-1} \left( \matr{x}(0) + \matr{B} \tilde{\matr{u}}[s] \right)
	\end{split}
	\end{equation}
	that gives the output in the frequency domain
	\begin{equation}
	\begin{split}
	\tilde{\matr{y}}[s] &=
	\matr{C} (s\matr{I} - \matr{A})^{-1} \matr{x}(0) \\
	&+ \matr{C} (s\matr{I} - \matr{A})^{-1} \matr{B} \tilde{\matr{u}}[s]
	+ \matr{D} \tilde{\matr{u}}[s]
	\end{split}
	\end{equation}
	Assuming that $\matr{x}(0) = \matr{0}$, the frequency domain input-output relation $\tilde{\matr{y}}[s] = \matr{G}[s] \tilde{\matr{u}}[s]$ is obtained for a transfer functions matrix
	\begin{equation}
	\matr{G}[s] = 
	\matr{C} (s\matr{I} - \matr{A})^{-1} \matr{B}
	+ \matr{D}
	\end{equation}
	
	\subsection{Transfer functions of the transducer}
			
		Signal flow graphs similar to Fig. \ref{fig:SignalFlowChart} can be drawn for each source in the equations of motion.
		Applying the rotating wave approximation, which is valid as long as the pump laser is set with a detuning $\Delta = \omega_m$ to the asymetric optical supermode, can be understood as neglecting the blocks of the creation operators in the signal flow graphs.
		
		In that limit, the annihilation operators transfer functions are given by:
		\begin{itemize}
			\item from the optical input $\hat{a}_{\text{in}}$ to the optical output $\hat{a}_{\text{out}}$
				\begin{equation}
					\begin{split}
						G_{ \hat{a}_{\mathrm{out}} \hat{a}_{\mathrm{in}}}[s] =\ 1
						- \frac{ \kappa_{\mathrm{ex}} \chi_{o, 1}[s] (1 + |g_{\mathrm{OM}}|^2 \chi_{o, 2}[s] \chi_{\mathrm{m}}[s]) }{ 1 + |g_{\mathrm{OM}}|^2 \chi_{o, 2}[s] \chi_{\mathrm{m}}[s] + J^2 \chi_{o, 1}[s] \chi_{o, 2}[s] }
					\end{split}
				\end{equation}
			\item from the microwave input $\hat{c}_{\text{in}}$ to the optical output $\hat{a}_{\text{out}}$
				\begin{equation}
					G_{ \hat{a}_{\mathrm{out}} \hat{c}_{\mathrm{in}}}[s] = 
					\frac{ \sqrt{\gamma_{\mathrm{ex}}} \sqrt{\kappa_{\mathrm{ex}}} g_{\mathrm{OM}} J \chi_{o, 1}[s] \chi_{o, 2}[s] \chi_{m}[s] }{ 1 + |g_{\mathrm{OM}}|^2 \chi_{o, 2}[s] \chi_{m}[s] + J^2 \chi_{o, 1}[s] \chi_{o, 2}[s] }
				\end{equation}
			\item from the noise in the first optical ring cavity $\hat{f}_{o, 1}$ to the optical output $\hat{a}_{\text{out}}$
				\begin{equation}
					G_{ \hat{a}_{\mathrm{out}} \hat{f}_{\mathrm{o}, 1} }[s] = 
					\frac{ - \sqrt{\kappa_{0, 1}} \sqrt{\kappa_{\mathrm{ex}}} \chi_{o, 1}[s] (1 + |g_{\mathrm{OM}}|^2 \chi_{o, 2}[s] \chi_{m}[s]) }{ 1 + |g_{\mathrm{OM}}|^2 \chi_{o, 2}[s] \chi_{m}[s] + J^2 \chi_{o, 1}[s] \chi_{o, 2}[s] }
				\end{equation}
			\item from the noise in the second optical ring cavity $\hat{f}_{o, 2}$ to the optical output $\hat{a}_{\text{out}}$
				\begin{equation}
					G_{ \hat{a}_{\mathrm{out}} \hat{f}_{\mathrm{o}, 2} }[s] = 
					\frac{ - i \sqrt{\kappa_{0, 2}} \sqrt{\kappa_{\mathrm{ex}}} \chi_{o, 1}[s] \chi_{o, 2}[s] J }{ 1 + |g_{\mathrm{OM}}|^2 \chi_{o, 2}[s] \chi_{m}[s] + J^2 \chi_{o, 1}[s] \chi_{o, 2}[s] }
				\end{equation}
			\item from the noise in the mechanical resonator $\hat{f}_{m}$ to the optical output $\hat{a}_{\text{out}}$
				\begin{equation}
					G_{ \hat{a}_{\mathrm{out}} \hat{f}_{m} }[s] = 
					\frac{ \sqrt{\gamma_{0}} \sqrt{\kappa_{\mathrm{ex}}} g_{\mathrm{OM}} J \chi_{o, 1}[s] \chi_{o, 2}[s] \chi_{m}[s] }{ 1 + |g_{\mathrm{OM}}|^2 \chi_{o, 2}[s] \chi_{m}[s] + J^2 \chi_{o, 1}[s] \chi_{o, 2}[s] }
				\end{equation}
		\end{itemize}	
		and for the microwave output $\hat{c}_{\text{out}}$
		\begin{itemize}
			\item from the optical input $\hat{a}_{\text{in}}$ to the microwave output $\hat{c}_{\text{out}}$
				\begin{equation}
					G_{ \hat{c}_{\mathrm{out}} \hat{a}_{\mathrm{in}}}[s] = 
					\frac{ - \sqrt{\gamma_{\mathrm{ex}}} \sqrt{\kappa_{\mathrm{ex}}} g_{\mathrm{OM}}^* J \chi_{o, 1}[s] \chi_{o, 2}[s] \chi_{m}[s] }{ 1 + |g_{\mathrm{OM}}|^2 \chi_{o, 2}[s] \chi_{m}[s] + J^2 \chi_{o, 1}[s] \chi_{o, 2}[s] }
				\end{equation}
			\item from the microwave input $\hat{c}_{\text{in}}$ to the microwave output $\hat{c}_{\text{out}}$
				\begin{equation}
					\begin{split}
						G_{ \hat{c}_{\mathrm{out}} \hat{c}_{\mathrm{in}}}[s] =\ -1
						+ \frac{ \gamma_{\mathrm{ex}} \chi_{m}[s] (1 + J^2 \chi_{o, 1}[s] \chi_{o, 2}[s] ) }{ 1 + |g_{\mathrm{OM}}|^2 \chi_{o, 2}[s] \chi_{\mathrm{m}}[s] + J^2 \chi_{o, 1}[s] \chi_{o, 2}[s] }
					\end{split}
				\end{equation}
			\item from the noise in the first optical ring cavity $\hat{f}_{o, 1}$ to the microwave output $\hat{c}_{\text{out}}$
				\begin{equation}
					G_{ \hat{c}_{\mathrm{out}} \hat{f}_{o, 1}}[s] = 
					\frac{ - \sqrt{\gamma_{\mathrm{ex}}} \sqrt{\kappa_{0, 1}} g_{\mathrm{OM}}^* J \chi_{o, 1}[s] \chi_{o, 2}[s] \chi_{m}[s] }{ 1 + |g_{\mathrm{OM}}|^2 \chi_{o, 2}[s] \chi_{m}[s] + J^2 \chi_{o, 1}[s] \chi_{o, 2}[s] }
				\end{equation}
			\item from the noise in the second optical ring cavity $\hat{f}_{o, 2}$ to the microwave output $\hat{c}_{\text{out}}$
				\begin{equation}
					G_{ \hat{c}_{\mathrm{out}} \hat{f}_{o, 2}}[s] = 
					\frac{ i \sqrt{\gamma_{\mathrm{ex}}} \sqrt{\kappa_{0, 2}} g_{\mathrm{OM}}^* \chi_{o, 2}[s] \chi_{m}[s] }{ 1 + |g_{\mathrm{OM}}|^2 \chi_{o, 2}[s] \chi_{m}[s] + J^2 \chi_{o, 1}[s] \chi_{o, 2}[s] }
				\end{equation}
			\item from the noise in the mechanical resonator $\hat{f}_{m}$ to the microwave output $\hat{c}_{\text{out}}$
				\begin{equation}
					\begin{split}
						G_{ \hat{c}_{\mathrm{out}} \hat{f}_{m}}[s] =\ 
						\frac{ \sqrt{\gamma_{0}} \sqrt{\gamma_{\mathrm{ex}}} \chi_{m}[s] (1 + J^2 \chi_{o, 1}[s] \chi_{o, 2}[s] ) }{ 1 + |g_{\mathrm{OM}}|^2 \chi_{o, 2}[s] \chi_{\mathrm{m}}[s] + J^2 \chi_{o, 1}[s] \chi_{o, 2}[s] }
					\end{split}
				\end{equation}
		\end{itemize}
	
	\subsection{Mason's gain of the microwave-to-optical transduction process}
	\label{section:TotalConversionEfficiency}
	
		The transfer function from the input microwave mode to the output optical mode can be obtained from Fig. \ref{fig:SignalFlowChart} by applying Mason's rule for the gain of signal flow graph \cite{mason1953feedback}
		\begin{equation}
		\label{eq:MasonGain}
		G = \frac{ \sum_k G_k \Delta_k }{\Delta}
		\end{equation}
		where the sum acts on all the possible forward paths from the input mode, the source, to the output mode, the sink.
		$G_k$ is the gain of the $k^{\text{th}}$ forward path, $\Delta_k$ is the subdeterminant for the loops not touching the $k^{\text{th}}$ forward path, and $\Delta = 1 - \sum_i L_i + \sum_{i, j}^{\text{not touching}} L_i L_j + \dots $ is the total determinant obtained from the individual loop gains $L_i$.
		The conversion efficiency of the transducer is obtained by first computing the transfer function $G$ between the annihilation operators $\hat{c}_{\text{in}}$ and $\hat{a}_{\text{out}}$, and then taking the square of its norm $|G|^2$.
		
		Here only two paths join $\hat{c}_{\text{in}}$ and $\hat{a}_{\text{out}}$ without going through the same nodes more than once:
		$\hat{c}_{\text{in}} \rightarrow \hat{b} \rightarrow \hat{a}_2 \rightarrow \hat{a}_1 \rightarrow \hat{a}_{\text{out}}$
		and $\hat{c}_{\text{in}} \rightarrow \hat{b} \rightarrow \hat{a}^{\dagger} \rightarrow \hat{b}^{\dagger} \rightarrow \hat{a}_2 \rightarrow \hat{a}_1 \rightarrow \hat{a}_{\text{out}}$.
		The authors would like to bring the attention of the reader to the fact that the self-loops of each mode have to be split in order to see the susceptibilities appearing.
		The gain of the first forward path is thus $\sqrt{\kappa_{\text{ex}}} \sqrt{\gamma_{\text{ex}}} J \chi_m[s] \chi_{o, 2}[s] \chi_{o, 1}[s] g_{\text{OM}}$, and the gain of the second forward path is $- \sqrt{\gamma_{\mathrm{ex}}} \sqrt{\kappa_{\mathrm{ex}}} \chi_{o, 1}[s] \chi_{o, 2}[s] \chi_m[s] g_{\mathrm{OM}} J  \chi_m^*[s^*] \chi_{o, 2}^*[s^*]  |g_{\mathrm{OM}}|^2 $.
		
		The next step to compute this transfer function is to identify all the loops of the graph.
		The loops with only one node are treated by considering that they are split in the following.
		No loop with more than two nodes is present in this graph.
		The loops to take into account are thus:
		\begin{itemize}
			\item $\hat{b} \leftrightarrow \hat{a}_{2} : L_{1} = - \chi_{o, 2}[s] \chi_m[s] |g_{\text{OM}}|^2$
			\item $\hat{b}^{\dagger} \leftrightarrow \hat{a}_{2}^{\dagger} : L_{2} = - \chi_{o, 2}^*[s^*] \chi_m^*[s^*] |g_{\text{OM}}|^2$
			\item $\hat{a}_{1} \leftrightarrow \hat{a}_{2} : L_{3} = - \chi_{o, 1}[s] \chi_{o, 2}[s] J^2$
			\item $\hat{a}_{1}^{\dagger} \leftrightarrow \hat{a}_{2}^{\dagger} : L_{4} = - \chi_{o, 1}^*[s^*] \chi_{o, 2}^*[s^*] J^2$
			\item $\hat{b} \leftrightarrow \hat{a}_{2}^{\dagger} : L_{5} = \chi_{o, 2}^*[s^*] \chi_m[s] |g_{\text{OM}}|^2$
			\item $\hat{b}^{\dagger} \leftrightarrow \hat{a}_{2} : L_{6} = \chi_{o, 2}[s] \chi_m^*[s^*] |g_{\text{OM}}|^2$
		\end{itemize}
		Only the loops $L_2$ and $L_4$ do not touch the first forward path, and they do overlap on the node $\hat{a}_2^{\dagger}$.
		The subdeterminant of the first forward path is thus
		\begin{align}
		\Delta_{\text{forward path 1}} = 
		1
		&+ \chi_{o, 2}^*[s^*] \chi_m^*[s^*] |g_{\text{OM}}|^2 \nonumber \\
		&+ \chi_{o, 1}^*[s^*] \chi_{o, 2}^*[s^*] J^2
		\end{align}
		The second forward path touches all the loops. Its subdeterminant is thus $\Delta_{\text{forward path 2}} = 1$
		
		The final step to obtain the total determinant is to identify the loops that are not touching:
		\begin{itemize}
			\item $L_{1}$ with $\left[ L_{2}, \  L_{4}\right]$
			\item $L_{2}$ with $\left[ L_{1}, \  L_{3}\right]$
			\item $L_{3}$ with $\left[ L_{2}, \  L_{4}, \  L_{5}\right]$
			\item $L_{4}$ with $\left[ L_{1}, \  L_{3}, \  L_{6}\right]$
			\item $L_{5}$ with $\left[ L_{3}, \  L_{6}\right]$
			\item $L_{6}$ with $\left[ L_{4}, \  L_{5}\right]$
		\end{itemize}
		
		The total determinant of the graph is thus
		\begin{widetext}
			\begin{equation}
				\begin{split}
					\Delta_{\text{total}} 
					=&\ 1 - L_{1} - L_{2} - L_{3} - L_{4} - L_{5} - L_{6} 
					+ L_{1} L_{2} + L_{1} L_{4} + L_{2} L_{3} + L_{3} L_{4} + L_{3} L_{5} + L_{4} L_{6} + L_{5} L_{6} \\
					=&\ 1 
					+( \chi_{o, 2}[s] \chi_m[s] |g_{\text{OM}}|^2)
					+( \chi_{o, 2}^*[s^*] \chi_m^*[s^*] |g_{\text{OM}}|^2)
					+( \chi_{o, 1}[s] \chi_{o, 2}[s] J^2)\\
					&+( \chi_{o, 1}^*[s^*] \chi_{o, 2}^*[s^*] J^2)
					-(\chi_{o, 2}^*[s^*] \chi_m[s] |g_{\text{OM}}|^2)
					-(\chi_{o, 2}[s] \chi_m^*[s^*] |g_{\text{OM}}|^2)\\ 
					&-( \chi_{o, 2}[s] \chi_m[s] |g_{\text{OM}}|^2) (- \chi_{o, 2}^*[s^*] \chi_m^*[s^*] | g_{\text{OM}}|^2)
					-( \chi_{o, 2}[s] \chi_m[s] |g_{\text{OM}}|^2) (- \chi_{o, 1}^*[s^*] \chi_{o, 2}^*[s^*] J^2)\\
					&-( \chi_{o, 2}^*[s^*] \chi_m^*[s^*] |g_{\text{OM}}|^2) (- \chi_{o, 1}[s] \chi_{o, 2}[s] J^2)
					-( \chi_{o, 1}[s] \chi_{o, 2}[s] J^2) (- \chi_{o, 1}^*[s^*] \chi_{o, 2}^*[s^*] J^2)\\
					&-( \chi_{o, 1}[s] \chi_{o, 2}[s] J^2) (\chi_{o, 2}^*[s^*] \chi_m[s] |g_{\text{OM}}|^2)
					-( \chi_{o, 1}^*[s^*] \chi_{o, 2}^*[s^*] J^2) (\chi_{o, 2}[s] \chi_m^*[s^*] |g_{\text{OM}}|^2)\\ 
					&+(\chi_{o, 2}^*[s^*] \chi_m[s] | g_{\text{OM}}|^2) (\chi_{o, 2}[s] \chi_m^*[s^*] | g_{\text{OM}}|^2)
				\end{split} 
			\end{equation}
		\end{widetext}
		Mason's rule (\ref{eq:MasonGain}) can now be applied to find the final result of section (\ref{section:TransductionTheory}), equations \eqref{eq:ConversionNumerator} and \eqref{eq:ConversionDenominator}.
			
	\subsection{Added noise}
	\label{section:AddedNoise}
		
\begin{figure*}
		\includegraphics[width=0.95\textwidth]{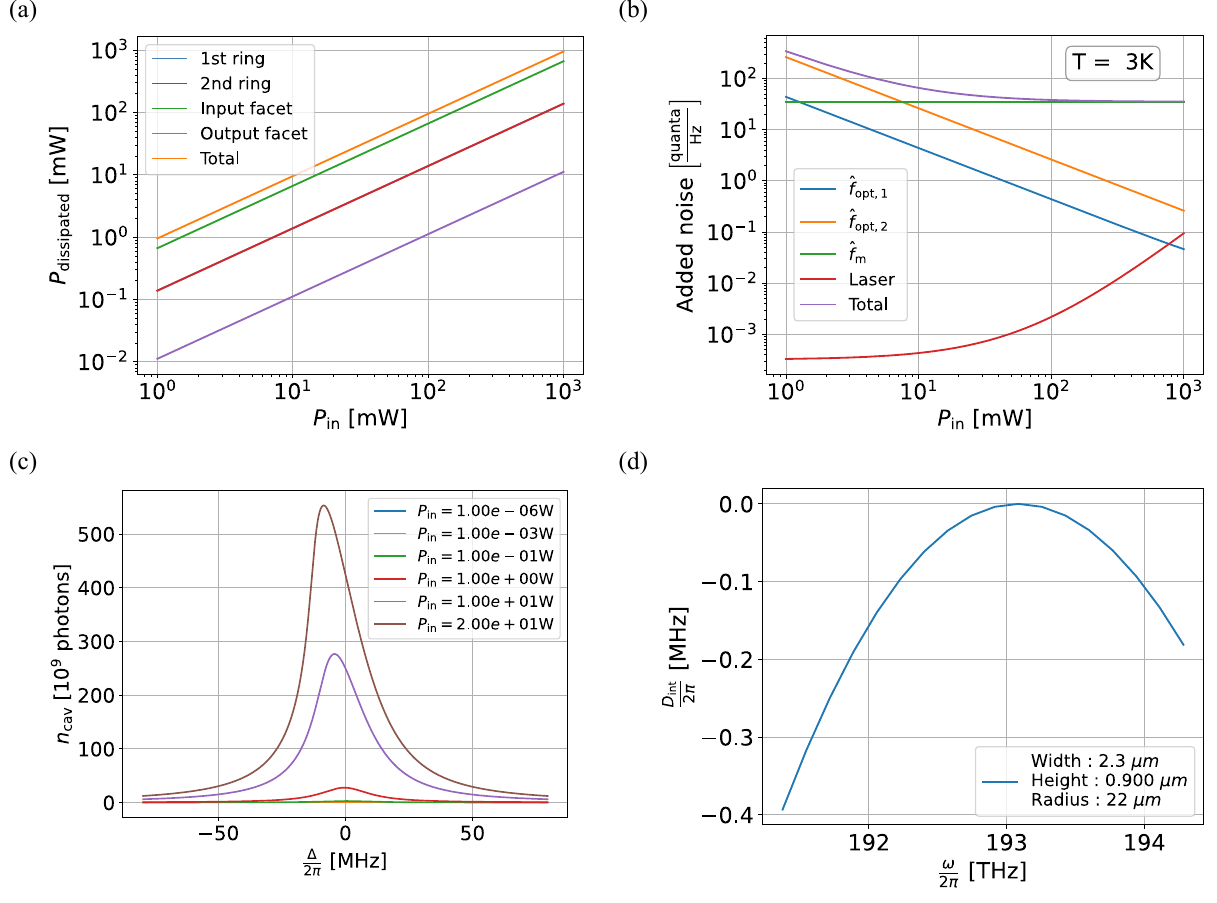}
		\caption{\textbf{Technical constraints on the transducer design.}
				(a) Heat load on the dilution fridge.
				Power dissipated while operating the transducer in the continuous wave regime, as a function of the on-chip power at the input of the trasnducer.
				(b) Added noise.
				Noise added during the conversion process for a transducer at the 3K stage of the fridge.
				(c) Optomechanical bistability.
				Intracavity photon number in a single \ch{Si3N4} optical ring cavity coupled to a HBAR resonance at $3.3$GHz as a function of the laser detuning.
				(d) Integrated dispersion.
				Dispersion curve of the TE optical mode in the silicon
				nitride waveguide. Negative values of the integrated dispersion insures
				to avoid Kerr parametric oscillations, which would prevent quantum
				coherent frequency conversion.}
		\label{fig:HeatLoad}
		\label{fig:AddedNoise}
		\label{fig:OM_bistability}
		\label{fig:IntegratedDispersion}
	\end{figure*}

		The optical output can be decomposed in the signal coming from the microwave and the noise added during the conversion process.
		\begin{align*}
			\delta \hat{a}_{\mathrm{out}}[\omega] &\approx 
			\sqrt{\eta} \left( \hat{c}_{\mathrm{in}}[\omega] + \sqrt{\frac{1-\eta}{\eta}} \hat{c}_{\mathrm{added}}[\omega] \right)
		\end{align*}
		Here the efficiency is given by the microwave-optical transfer function expressed in terms of quanta $\eta = |G_{ \hat{a}_{\mathrm{out}} \hat{c}_{\mathrm{in}}}[\omega]|^2$, while the added noise $\sqrt{\frac{1-\eta}{\eta}} \hat{c}_{\mathrm{added}}[\omega]$ is related to the transfer functions of the other sources and their power spectral density.
		\begin{align*}
			\sqrt{ 1-\eta } \ \hat{c}_{\mathrm{added}}[\omega] &=
			G_{\hat{a}_{\text{out}} \hat{a}_{\text{in}} }[\omega] \delta \hat{a}_{\mathrm{in}}[\omega]
			+ G_{\hat{a}_{\text{out}} \hat{f}_{\mathrm{o}, 1} }[\omega] \hat{f}_{\mathrm{o}, 1}[\omega] \\
			&+ G_{\hat{a}_{\text{out}} \hat{f}_{\mathrm{o}, 2} }[\omega] \hat{f}_{\mathrm{o}, 2}[\omega]
			+ G_{\hat{a}_{\text{out}} \hat{f}_{\mathrm{MW}} }[\omega] \hat{f}_{\mathrm{MW}}[\omega]
		\end{align*}
		Here the first term corresponds to the laser noise, the second and the third terms correspond to the thermal noise in the optical ring cavities, and the fourth one corresponds to the thermal noise from the HBAR.
		Each noise term thus contributes to $\left| \frac{ G_{\hat{a}_{\text{out}} \hat{f} }[\omega] }{ G_{ \hat{a}_{\mathrm{out}} \hat{b}_{\mathrm{in}}}[\omega] } \right|^2
		S_{\hat{f} \hat{f} }[\omega]$ added noise quanta to the output, where $S_{\hat{f} \hat{f}}[\omega] = \frac{1}{2} \left(\expval{\hat{f}[\omega] \hat{f}^{\dagger}[\omega]} + \expval{\hat{f}^{\dagger}[\omega] \hat{f}[\omega]} \right)$ refers to the symmetrized power spectral density \cite{clerk_introduction_2010}.
		The high pump powers required to reach sufficient conversion efficiencies imply that the transducer has to be operated on the 3K flange of the fridge, which provides sufficient cooling power.
		Fig. \ref{fig:AddedNoise}\textcolor{NavyBlue}{b}) shows the noise added at the optical output for a transducer at the 3K flange of the fridge, using  Bose-Einstein distribution $n_{\mathrm{th}}[\omega, T] = \frac{1}{e^{\frac{\hbar\omega}{k_B T}} - 1}$ for the thermal occupation in each mode.
		The noise from the term $G_{\hat{a}_{\text{out}} \hat{a}_{\text{in}} }[\omega] \delta \hat{a}_{\mathrm{in}}[\omega]$ correspond to the noise coming from the pump laser. Given that the optomechanical interaction depends on the laser intensity and not its phase, the only noise to take into account here is the relative intensity noise.
		Assuming a value of \SI{1e-14}{\per\hertz} for the RIN \cite{lecocq_control_2021} and taking into account the line shape of the laser, the noise transducer from the pump laser is found to be much smaller than a quantum.
		
		The added noise is dominated by the contribution of the thermal excitations in the acoustic mode, the optical noise only contributing to a few quanta of added noise.
		While direct conversion is not quantum limited, the low noise level and high conversion efficiency would enable to use protocols for perfect transduction \cite{Lau2019, wu_deterministic_2021}.
		
\section{Technical limitations}

	\subsection{Heat load on the dilution fridge}
	
		As consequent power is required to reach the maximal efficiency, it is important to verify that the dilution fridge will be able to provide enough cooling power to maintain the $3K$ environment.
		We identify two sources of heat from the transducer: the power dissipated on the chip, and the power scattered from the fiber-chip interfaces.
		In order to get an upper bound on the heat load, we assumed that the entire dissipated on chip and scattered from the chip facets is converted into heat.
		The fiber-chip insertion loss is approximatively $40\%$ per facet \cite{liu2020high}, implying that roughly $60\%$ of the power on the fiber reaches the transducer.
		The power at the output facet is reduced because most of the photons are absorbed in the cavity.
		The power dissipated inside the transducer estimated assuming the intracavity optical photons are converted into heat at a rate given by the intrinsic optical linewidth. The two optical ring cavities thus give contributions of $\hbar \omega_1 \kappa_{0, 1} |\bar{a}_1|^2$ and $\hbar \omega_2 \kappa_{0, 2} |\bar{a}_2|^2$ respectively.

		Fig. \ref{fig:HeatLoad}\textcolor{NavyBlue}{a}) gives an estimation of the heat load, based on the formula
		\begin{align*}
			P_{\mathrm{dissipated}} =\ & 
			\frac{ \mathrm{IL} }{ 1 - \mathrm{IL} } P_{\mathrm{in}}
			+ \mathrm{IL} |G_{ a_{\mathrm{in}} a_{\mathrm{out}} }|^2 P_{\mathrm{in}} \\
			&+ \hbar \omega_1 \kappa_{0, 1} |\bar{a}_1|^2
			+ \hbar \omega_2 \kappa_{0, 2} |\bar{a}_2|^2
		\end{align*}
		We conclude that the dilution fridge has to compensate for heat corresponding of the order of magnitude of the on-chip power at the input of the transducer. While this load is surpasses the limit of of the \si{\milli\kelvin} stages, the \SI{3}{\kelvin} flange can provide cooling power $> \SI{1}{\watt}$. The transducer can thus be operated on that stage.

	\subsection{Kerr parametric instabilities}
	
		As silicon nitride microrings exhibit ultra-low propagation losses together with decent Kerr coefficients, they are known for the generation of Kerr frequency combs with low pump laser powers. Unfortunately this effect would here constitute an additional loss channel, and must therefore be avoided.
		We refer to \cite{Herr2012}, that indicates that the first sidebands occurs at the mode number
		\begin{equation}
			\mu_{\mathrm{thresh}} = \sqrt{ \frac{\kappa}{D_2} \left( \sqrt{ \frac{ P_{\mathrm{in}} }{ P_{\mathrm{thresh}} }  - 1 } - 1 \right) }
		\end{equation}
		This equation implies that Kerr parametric instabilities will not occur for normal dispersion $\frac{d^2\omega}{dk^2}<0$.
		Further studies have shown that for the parametric oscillations to occur in the case of normal dispersion, the laser has to be detuned from the resonance \cite{PhysRevA.89.063814}, which is not the case in the scheme presented here.
		In practice, modes crossing might still compensate for the normal dispersion and there is a small chance that high pump power produce this deleterious effect. In this case, changing the wavelength of operation will allow to avoid the local anomalous dispersion.
		Nonetheless, it is important to make sure that the integrated dispersion is as far as possible from the anomalous regime.
		The Fig. \ref{fig:IntegratedDispersion}\textcolor{NavyBlue}{d}) shows weak normal dispersion for the chosen set of parameters, which were optimized relatively to other constrains, such as for example the fact that the effective mass of the HBAR mode increases with the ring radius.
	
	\subsection{Optomechanical bistability}
				
		Strong optomechanical coupling can bring the optical cavity to a bistable regime, where two different intracavity photon numbers are stable solutions for the steady state of the optomechanical system \cite{bowen2015quantum}.
		This process would limit the pump laser power that could be use to operate the transducer.
		Fig. \ref{fig:OM_bistability}\textcolor{NavyBlue}{c}) shows a simulation of the intracavity photon number for a single \ch{Si3N4} optical ring cavity,
		\begin{equation}
			\left[ \left(\Delta + \frac{2 g_0^2 n_{\mathrm{cav}}}{\omega_m}\right)^2 + \frac{\kappa^2}{4} \right] n_{\mathrm{cav}} = \kappa_{\mathrm{ex}} \frac{P_{\mathrm{in}}}{\hbar\omega_L}
		\end{equation}
		which would indicate the upper limit on the power that can be send in the coupled rings configuration.
		This figure shows that the bistable regime cannot be reached with the system, even for powers larger than \SI{1}{\watt}.
	

\bibliography{Bibliography}{}

\end{document}